\documentclass[preprintnumbers,amsmath,amssymb,superscriptaddress]{revtex4}
\usepackage{tikz}
\usepackage{bm}
\usepackage{graphicx}
\usepackage{chemarrow}
\usepackage{multirow}

\begin{document}

%----------------------------------------------------------------------------------------------
\title{Mirror symmetry breaking with limited
enantioselective autocatalysis and temperature gradients: a
stability survey}
%----------------------------------------------------------------------------------------------

\author{Celia Blanco}
\email{blancodtc@cab.inta-csic.es} \affiliation{Centro de
Astrobiolog\'{\i}a (CSIC-INTA), Carretera Ajalvir Kil\'{o}metro 4,
28850 Torrej\'{o}n de Ardoz, Madrid, Spain}

\author{Josep M. Rib\'{o}}
\email{jmribo@ub.es} \affiliation{Department of Organic Chemistry,
University of Barcelona, c. Mart\'{i} i Franqu\`{e}s 1, 08028
Barcelona, Catalonia, Spain.} \affiliation{Institute of Cosmos
Science, University of Barcelona (IEEC-UB), c. Mart\'{i} i
Franqu\`{e}s 1, 08028 Barcelona, Catalonia, Spain.}

\author{Joaquim Crusats}
\email{j.crusats@ub.es} \affiliation{Department of Organic
Chemistry, University of Barcelona, c. Mart\'{i} i Franqu\`{e}s 1,
08028 Barcelona, Catalonia, Spain.}\affiliation{Institute of Cosmos
Science, University of Barcelona (IEEC-UB), c. Mart\'{i} i
Franqu\`{e}s 1, 08028 Barcelona, Catalonia, Spain.}

\author{Zoubir El-Hachemi}
\email{zelhachemi@ub.es} \affiliation{Department of Organic
Chemistry, University of Barcelona, c. Mart\'{i} i Franqu\`{e}s 1,
08028 Barcelona, Catalonia, Spain.}\affiliation{Institute of Cosmos
Science, University of Barcelona (IEEC-UB), c. Mart\'{i} i
Franqu\`{e}s 1, 08028 Barcelona, Catalonia, Spain.}

\author{Albert Moyano}
\email{amoyano@ub.es} \affiliation{Department of Organic Chemistry,
University of Barcelona, c. Mart\'{i} i Franqu\`{e}s 1, 08028
Barcelona, Catalonia, Spain.}

\author{David Hochberg}
\email{hochbergd@cab.inta-csic.es} \affiliation{Centro de
Astrobiolog\'{\i}a (CSIC-INTA), Carretera Ajalvir Kil\'{o}metro 4,
28850 Torrej\'{o}n de Ardoz, Madrid, Spain}

\begin{abstract}
We analyze limited enantioselective (LES) autocatalysis in a
temperature gradient and with internal flow/recycling of hot and
cold material.  Microreversibility forbids broken mirror symmetry
for LES in the presence of a temperature gradient alone. This
symmetry can be broken however when the auto-catalysis and limited
enantioselective catalysis are each localized within the regions of
low and high temperature, respectively. This scheme has been
recently proposed as a plausible model for spontaneous emergence of
chirality in abyssal hydrothermal vents. Regions in chemical
parameter space are mapped out in which the racemic state is
unstable and bifurcates to chiral solutions.
\end{abstract}

\maketitle

%----------------------------------------------------------------
\section{\label{sec:intro} Introduction}
%---------------------------------------------------------------

Recent experimental reports on the deracemization of racemic
mixtures of crystals and on the crystallization from boiling
solutions
\cite{CV2011,El-Hach2011,Noorduin2008,Noorduin2009,Viedma2005,
Wattis2011} are striking examples of novel scenarios for spontaneous
mirror symmetry breaking (SMSB) for compounds for which the
homochiral interactions are favored over the heterochiral ones. In
other words, these are reactions that cannot be explained by
Frank-like mechanisms, in which the heterochiral interaction is the
favored one. Despite some controversy about the actual mechanisms
responsible for the SMSB in these situations, the experimental
reports all coincide in that the final state is stationary:
mechano-stationary in the case of wet grinding of racemic mixtures
of crystals \cite{Noorduin2008,Noorduin2009,Viedma2005} and the
presence of temperature gradients in the case of deracemization and
crystallization in boiling solutions \cite{CV2011,El-Hach2011}. The
above reports stress the fact that the racemic conglomerate crystal
mixtures are deracemized in experimental conditions where
\textit{chemical equilibrium is not possible}, i.e., they require
specific energy input to only some of the species of the system (as
in the crystal grinding experiments) or else a non-uniform
temperature distribution.

The point of departure for the present paper is: if limited
enantioselectivity \cite{AG} in experimental conditions of closed
systems with a uniform distribution of temperature and energy
inexorably  yields a final racemic state, \cite{RH} then can SMSB
occur in scenarios of non-uniform temperature distributions?  As it
turns out, this is in fact possible in the case that, in addition to
a non-uniform temperature distribution, the systems possess a
compartmentalization of the enantioselective and the
non-enantioselective autocatalyses.

The emergence of chirality in enantioselective autocatalysis for
compounds which do not follow Frank-like schemes is investigated
here for the limited enantioselectivity (LES) model composed of
coupled enantioselective and non-enantioselective autocatalyses. The
basic model \cite{AG} is defined by the following chemical
transformations.
\noindent Production of chiral compounds L,D from an achiral
substrate A:
\begin{eqnarray}\label{decay}
\textrm{A} \autorightleftharpoons{$k_1$}{$k_{-1}$} \textrm{L},
\qquad \textrm{A} \autorightleftharpoons{$k_1$}{$k_{-1}$}
\textrm{D}.
\end{eqnarray}

\noindent Autocatalytic production:
\begin{eqnarray}\label{autoLD2}
\textrm{A} + \textrm{L} \autorightleftharpoons{$k_2$}{$k_{-2}$}
\textrm{L} + \textrm{L}, \qquad \textrm{A} + \textrm{D}
\autorightleftharpoons{$k_2$}{$k_{-2}$} \textrm{D} + \textrm{D}.
\end{eqnarray}

\noindent Limited enantioselectivity:
\begin{eqnarray}\label{limenant}
\textrm{A} + \textrm{L} \autorightleftharpoons{$k_3$}{$k_{-3}$}
\textrm{L} + \textrm{D}, \qquad \textrm{A} + \textrm{D}
\autorightleftharpoons{$k_3$}{$k_{-3}$} \textrm{D} + \textrm{L}.
\end{eqnarray}

In contrast to the Frank model, LES is able to account for two
important facts: namely, (i) the enantioselectivity of any chiral
catalyst is limited because of the third reaction Eq.(\ref{limenant}
), and (ii) the kinetic link between mirror conjugate processes
arises from the reversibility of the catalytic stage \cite{AG}. The
inverse reaction of the non-enantioselective autocatalysis (reaction
(\ref{limenant})) substitutes for the mutual inhibition reaction in
the Frank model or formation of the heterodimer (L + D $\rightarrow$
P). Earlier reports had claimed spontaneous mirror symmetry breaking
(SMSB) in LES, but this cannot occur in either open or closed
systems with a \textit{uniform} temperature distribution. The
obstacle comes from microreversibility, where $K(T)$ is the
temperature dependent equilibrium constant:
\begin{equation}\label{microrev1}
\frac{k_i}{k_{-i}} = K(T), \qquad (1 \leq i \leq 3).
\end{equation}
The condition for the instability of the racemic state is that
\begin{equation}\label{ineq1}
0 < g < \frac{1 - w}{1+3w} < 1,
\end{equation}
where $g = \frac{k_{-2}}{k_{-3}}$ and $w = \frac{k_3}{k_2}$
\cite{RH}. From Eq. (\ref{ineq1}), we must have $1-w > 0$ so that $1
> w$. But from Eq. (\ref{microrev1}), $\frac{k_2}{k_3} =
\frac{k_{-2}}{k_{-3}}$, which is incompatible with $1>w$ and $g<1$.
This is the situation for open systems. For closed systems, it can
be shown \cite{RH} that the racemic state is unstable provided that
\begin{equation}\label{closedLES}
0 < g < g_{crit}^{closed} < g_{crit}^{open} = \frac{1 - w}{1+3w} <
1,
\end{equation}
where the critical parameter in a closed system is always bounded
above by the corresponding one for open systems, and approaches the
latter from below in the limit of large total concentrations $C$,
that is: $g_{crit}^{closed} \rightarrow g_{crit}^{open}$. But  Eq.
(\ref{closedLES}) is also incompatible with Eq. (\ref{microrev1}).
So the racemic state is always asymptotically stable in this scheme
for both open and closed systems held at a uniform temperature.
Therefore there can be no asymptotically stable chiral outcome in
this model.

Nor can the LES model lead to SMSB in closed systems even with a
stationary non-uniform temperature distribution. On the other hand,
numerical simulations of chemical kinetics in a two-compartment
model \cite{vents} demonstrate that SMSB may occur if both catalytic
reactions (\ref{autoLD2}) and (\ref{limenant}) are spatially
separated at different temperatures in different compartments but
coupled under the action of a continous internal flow. In such
conditions the system can evolve, for certain reaction and system
parameters, towards a chiral stationary state, i.e., the system is
able to reach a bifurcation point leading to SMSB.

This is an appealing result since numerical simulations using
reasonable chemical parameters suggest that an adequate scenario for
such an LES-based SMSB would be that of abyssal hydrothermal vents,
by virtue of the typical temperature gradients found there and the
role of inorganic solids mediating chemical reactions in an
enzyme-like role. We therefore proposed \cite{vents} that a natural
prebiotic scenario for such emergence of chirality is that of
abyssal hydrothermal vents and volcanic plumes
\cite{Baross,Holm,Martin,Miller} which do have the adequate
temperature gradients and contain solids, as for example clays,
which have been proposed by several authors
\cite{Bernal,Brack,Cairns,Ferris,Hazen} as catalysts in the
prebiotic synthesis of organic compounds.

In view of the above, this paper deals with an analytic/numerical
study of the conditions leading to the instability of the ideal
racemic composition for the LES model with compartmentalized
catalysis (\ref{autoLD2}) and (\ref{limenant}) in regions held at
different temperatures. The two-compartment model (Sec
\ref{sec:LES-Tgrad}) is already sufficiently involved as to make
deriving general analytic stability results a near impossible task.
We thus focus our efforts on analyzing properties of the racemic
fixed point; the analytic conditions for its linear stability can be
set up and then tested in numerical domains. The direct study of the
stationary chiral solutions is substantially more complicated and
algebraically unwieldy. Nevertheless, the characterization of a
racemic state as unstable necessarily implies that the system
evolves to a state of non-racemic composition. Such a nonracemic
state could be stable, chaotic or oscillatory. However, the many
numerical SMSB tests performed for reasonable reaction parameters
and under the thermodynamic constraints imposed by the principle of
microreversibility and the temperature dependencies of the reaction
rate constants (see Sec \ref{sec:arrhenius}) have in all cases led
to a stable chiral state. These stable states are scalemic mixtures
of enantiomeric excesses (ee) whose values depend on the parameters
of the phase diagram. We first review the impossibility of SMSB in
LES with a stable temperature gradient. We then prove that SMSB in
such a system is possible when the two catalyses (\ref{autoLD2}) and
(\ref{limenant}) are compartmentalized (localized) in different
temperature regions connected by an internal flow of material.

%------------------------------------------------------------
\section{\label{sec:LES-Tgrad} LES with a temperature gradient}
%------------------------------------------------------------

\begin{figure}[h]
\centering
\includegraphics[width=0.45\textwidth]{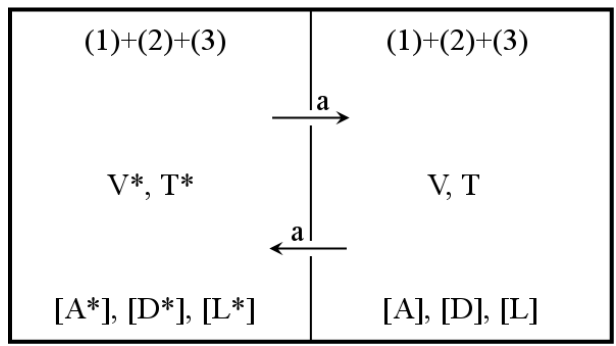}
\caption{\label{twocomps} Limited enantioselectivity (LES) in two
compartments with volumns $V$ and $V^*$ each held at different
temperatures $T^*
> T$ and interconnected by an internal flow $a$ of material. See
Ref. \cite{vents}.}
\end{figure}

Consider the LES scheme in a temperature gradient: this permits the
reaction rates to vary spatially in the system, from one place to
another, and might provide a way to achieve mirror symmetry
breaking. The reverse reaction of (\ref{limenant}) in one region
(compartment) could be faster that the reverse of (\ref{autoLD2}) in
the other. And then mixing could bring the hot and cold material
into contact. We model this by a closed two-compartment system with
volumes $V,V^*$; each compartment held at a uniform temperature,
$T^* > T$, and internally coupled by a constant internal flow or
recycling. The concentrations, and reaction rates for the second
compartment are labeled by an asterisk, see Figure \ref{twocomps}.
We extend the basic LES scheme
Eqs.(\ref{decay},\ref{autoLD2},\ref{limenant}), by including the
terms corresponding to the internal flow. The internal ``flow"
parameter $a$ has units of volume/time; $V$ and $V^*$ denote the
volumes of each compartment. The corresponding rate equations for
the two-compartment system have been derived in \cite{vents}.

%--------------------------------------------------------
\subsection{\label{sec:newvariables} New variables}
%-------------------------------------------------------

For setting up a stability analysis it is convenient to employ the
sums and differences of the concentrations $\chi=[L]+[D]$, $y
=[L]-[D]$, $\chi^*=[L^*]+[D^*]$ and $y^*=[L^*]-[D^*]$. In terms of
these, the rate equations take the following form:
\begin{eqnarray}\label{dAd}
\frac{d [A]}{dt} &=& - 2k_1[A]-(k_2[A]+k_3[A]-k_{-1})\chi +
\frac{k_{-2}}{2}(y^2 + \chi^2) + \frac{k_{-3}}{2}(\chi^2 -
y^2)+\frac{a}{V}([A^*]-[A]),
\end{eqnarray}

\begin{eqnarray}\label{dchid}
\frac{d \chi}{dt} &=& 2k_1[A]+(k_2[A]-k_{-1})\chi+k_3[A]\chi
-k_{-2}\frac{(\chi^2 + y^2)}{2}-k_{-3}\frac{(\chi^2-y^2)}{2}
+\frac{a}{V}(\chi^*-\chi),
\end{eqnarray}

\begin{eqnarray}\label{detad}
\frac{d y}{dt} &=& (k_2[A]-k_{-1})y -k_{-2}\chi y - k_3[A]y
+\frac{a}{V}(y^*-y),
\end{eqnarray}

\begin{eqnarray}\label{dAl}
\frac{d [A^*]}{dt} &=& -
2k_1^*[A^*]-(k_2^*[A^*]+k_3^*[A^*]-k_{-1}^*)\chi^* +
k_{-2}^*({y^*}^2 + {\chi^*}^2)+
\frac{k_{-3}^*}{2}({\chi^*}^2-{y^*}^2) +\frac{a}{V^*}([A]-[A^*]),
\end{eqnarray}

\begin{eqnarray}\label{dchil}
\frac{d \chi^*}{dt} &=&
2k_1^*[A^*]+(k_2^*[A^*]-k_{-1}^*)\chi^*+k_3^*[A^*]\chi^*
-k_{-2}^*\frac{(\chi^{*2}+y^{*2})}{2}-k_{-3}^*\frac{(\chi^{*2}-y^{*2})}{2}+\frac{a}{V^*}(\chi-\chi^*),
\end{eqnarray}

\begin{eqnarray}\label{detal}
\frac{d y^*}{dt}&=& (k_2^*[A^*]-k_{-1}^*)y^*-k_{-2}^*\chi^*y^*
-k_3^*[A^*]y^* +\frac{a}{V^*}(y-y^*).
\end{eqnarray}

They satisfy the constant mass constraint:
\begin{equation}\label{massconst}
V([A]+\chi)+V^*([A^*]+\chi^*)=C,
\end{equation}
where $C$ is the total conserved mass in the complete
two-compartment system.

%------------------------------------------------------------------------
\section{\label{sec:arrhenius} Constraints from Arrhenius-Eyring}
%-----------------------------------------------------------------------

Certain thermodynamic relationships hold among the reaction rates in
both compartments. These will be used to prove that SMSB is also
impossible for the scheme presented in Sec \ref{sec:LES-Tgrad} in a
background temperature gradient. Following this demonstration, we
then introduce the variant of LES that can and does lead to SMSB.

From Arrhenius-Eyring, the forward (and reverse) reaction rates for
reaction $i$ at temperatures $T^*$ and $T < T^*$ are \cite{Chang}
\begin{eqnarray}
k_i^* &=& \big( \frac{k_B T^*}{h} \big)\, e^{-\frac{\Delta
G_i(T^*)}{RT^*}}, \qquad k_{-i}^* = \big( \frac{k_B T^*}{h} \big)\,
e^{-\frac{\Delta
G_{-i}(T^*)}{RT^*}}, \nonumber\\\\
k_i &=& \big( \frac{k_B T}{h} \big) \, e^{-\frac{\Delta
G_i(T)}{RT}}, \qquad k_{-i} = \big( \frac{k_B T}{h} \big) \,
e^{-\frac{\Delta G_{-i}(T)}{RT}},\nonumber\\
\end{eqnarray}
here
\begin{equation}\label{deltafreeforward}
\Delta G_i(T) = \Delta H_i - T\Delta S_i
\end{equation}
denotes the difference in free energy between the activated state
(transition state) and the reactants, while
\begin{equation}\label{deltafreereverse}
\Delta G_{-i}(T) = \Delta H_{-i} - T\Delta S_{-i},
\end{equation}
is the free energy difference between the activated state
(transition state) and the products; $H$ and $S$ denote the enthalpy
and entropy, respectively. From the above we can obtain a relation
between the forward reaction rates $i$ in each compartment:
\begin{equation}
k^*_i = \big( \frac{T^*}{T} \big) \, \exp \Big( -\frac{\Delta
H_i}{R}(\frac{1}{T^*}-\frac{1}{T}) \Big) k_i.
\end{equation}
Clearly, once the values of the $k_i$ are chosen for the reference
compartment at $T$, we are not free to \textit{independently} choose
the reaction rates $k^*_i$ at the higher temperature $T^*$.

The fundamental microreversibility condition Eq.(\ref{microrev1})
together with Arrhenius-Eyring implies
\begin{eqnarray}
\frac{k_i}{k_{-i}} &=& \frac{e^{-\frac{\Delta G_i}{RT}}}{
e^{-\frac{\Delta G_{-i}}{RT}}} = e^{(\Delta G_{-i}-\Delta G_i)/RT} =
K(T),\\
\nonumber \\
 &\Leftrightarrow& \Delta G_{-i}-\Delta G_i \equiv \Delta \Delta
 G,
\end{eqnarray}
that is, the difference of the free energy differences $\Delta
\Delta G$ must be \textit{independent} of $i$, that is, independent
of the specific $ith$ reaction. This implies that the individual
double differences in enthalpy and in entropy must also be
independent of reaction $i$, so we must also have
\begin{eqnarray}
\Delta \Delta H = (\Delta H_{-i} - \Delta H_{i}),\\
\Delta \Delta S = (\Delta S_{-i} - \Delta S_{i}).
\end{eqnarray}
This also gives us an expression for calculating $K$:
\begin{equation}\label{KT}
K(T) = e^{\frac{\Delta \Delta H}{RT}}e^{-\frac{\Delta \Delta S}{R}}.
\end{equation}
The inverse reaction rates are obtained through the constraint
Eq.(\ref{microrev1}) as follows:
\begin{equation}\label{inverse}
k_{-i} = \frac{k_i}{K(T)}.
\end{equation}

We also note that if the constraint Eq. (\ref{microrev1}) is
satisfied at one specific temperature $T$, then it will
automatically hold at all others, that is
\begin{equation}\label{microrev*}
\frac{k^*_i}{k^*_{-i}} = K(T^*) = e^{\frac{\Delta \Delta
H}{RT^*}}e^{-\frac{\Delta \Delta S}{R}}, \qquad (1 \leq i \leq 3).
\end{equation}
The ratio of the equilibrium constants is given by
\begin{equation}\label{ratioK}
\frac{K(T^*)}{K(T)} =\exp\Big( \frac{\Delta \Delta
H}{R}(\frac{1}{T^*}-\frac{1}{T}) \Big).
\end{equation}
The algebraic intricacy of the model in Sec \ref{sec:LES-Tgrad} is
already such that we are unable to obtain \textit{useful} and
manageable analytic closed form expressions for the conditions
leading to the instability of the racemic solution. The situation is
even worse for obtaining analytic information regarding the possible
stationary chiral solutions. We appeal instead to chemically
inspired conjectures that can be tested numerically for coherence
and compatibility with microreversibility.

First, it is clear that in view of the gradient $T<T^*$, the
putative condition, which could conceivably lead to symmetry
breaking in the limit of small values of $a$,
\begin{eqnarray}\label{conjecture1}
&&k_{-2} < k_{-3} \qquad \& \qquad k_2 > k_3 \qquad {\rm at}\quad T
\qquad \nonumber\\
{\rm and}  &&k_{-2}^* < k^*_{-3} \qquad \& \qquad k_2^*
> k_3^* \qquad {\rm at}\quad T^*,\nonumber\\
\end{eqnarray}
is incompatible with the constraints in Eqs.
(\ref{KT},\ref{microrev*}). This condition (\ref{conjecture1}) is
inspired by the observation that for $a \rightarrow 0$, the two
compartments are practically isolated from each other and can be
treated as approximately independent. These are thus the necessary
conditions for obtaining an unstable racemic solution in each
compartment (see Sec \ref{sec:intro}). But they are incompatible
with microreversibility.

Secondly, the analysis in Sec \ref{sec:intro} suggests that SMSB
might occur when the inverse reaction of (\ref{limenant}) in one
region is \textit{faster} than the inverse reaction of
(\ref{autoLD2}) in the other region. Taking microreversibility into
account, the only way this might be achieved is, for example, by
arranging for
\begin{eqnarray}
&&k_{-3} < k_{-2} \qquad \& \qquad k_2 > k_3 \qquad {\rm at}\quad T
\qquad \nonumber\\
{\rm and}  &&k_{-2}^* < k^*_{-3} \qquad \& \qquad k_2^* <
k_3^* \qquad {\rm at}\quad T^*.\nonumber\\
\end{eqnarray}
But this is forbidden by virtue of Eq. (\ref{ratioK}), which is
satisfied by the ratio of the equilibrium constants. So a
temperature gradient and internal flow are by themselves, not enough
to produce a bifurcation. Actually, no spatially varying temperature
profile is sufficient, as can be seen by partitioning the closed
system into a number of sufficiently small regions within which the
local temperature is approximately uniform.

%---------------------------------------------------------------
\section{Temperature gradient and immobilized catalysts}
%---------------------------------------------------------------

Our working hypothesis is that a \textit{necessary but not
sufficient} condition for the instability of the racemic solution is
$k^*_{-3}
> k_{-2}$ and $k_2 > k^*_3$ in the presence of immobilized catalysts that
ensure that $k^*_{\pm 2} = 0$ in one region and $k_{\pm 3} = 0$ in
the other \cite{vents}.  Other conditions (total system
concentration $C$, the flow rate $a$, compartment volumes $V,V^*$,
etc.) also come into play for determining the overall instability,
in a highly nontrivial and nonlinear fashion.

%---------------------------------------------------------------
\subsection{Linear stability analysis of the stationary racemic fixed point}
%--------------------------------------------------------------

The equations (\ref{detad},\ref{detal}) for $\frac{d y}{dt} =
\frac{d y^*}{dt} = 0$ are identically satisfied for the stationary
solution $\bar y = \bar y^* = 0$. We therefore carry out a stability
analysis of this racemic fixed point and determine whether the
racemic solution (racemic in \textit{both} compartments) is
asymptotically stable or unstable. This will depend on the internal
\textit{flow parameter} $a$ that characterizes the cycling of hot to
cold material between the two compartments. Clearly, if we set
$a=0$, we merely recover two isolated copies of LES in independent
closed compartments, each one at a constant temperature, and there
can be no mirror symmetry breaking in this situation; the
considerations of Sec \ref{sec:intro} apply.

An algebraic advantage of studying the racemic fixed point $\bar y =
\bar y^* = 0$ is that the five independent concentration
fluctuations decouple into two sets of three and two, respectively.
This situation is reflected in the structure of the Jacobian $J$
which then reduces to a block-diagonal form with a $3\times3$
sub-block corresponding to the fluctuations $(\delta A, \delta \chi,
\delta \chi^*)$ and a $2\times2$ sub-block corresponding to $(\delta
y, \delta y^*)$ thus:
\begin{equation}\label{Jacobian1}
J = \left(
                \begin{array}{cc}
                  A^{3\times3} & D^{3\times2} \\
                  C^{2\times3} & B^{2\times2}  \\
                \end{array}
              \right) \Rightarrow \left(
                \begin{array}{cc}
                  A^{3\times3} & 0 \\
                  0 & B^{2\times2} \\
                \end{array}
              \right).
\end{equation}
The temporal evolution of the linearized concentration fluctuations
about the racemic fixed point $\bar y = \bar y^* = 0$ of the kinetic
equations is given by

\begin{eqnarray}
&&\frac{d}{dt} \left(
               \begin{array}{c}
                 \delta A \\
                 \delta \chi \\
                 \delta \chi^* \\
               \end{array}
             \right) = A \left(
               \begin{array}{c}
                 \delta A \\
                 \delta \chi \\
                 \delta \chi^* \\
               \end{array}\right) \nonumber\\
{\rm and} &&\frac{d}{dt} \left(
               \begin{array}{c}
                 \delta y \\
                 \delta y^* \\
               \end{array}
             \right) = B \left(
               \begin{array}{c}
                 \delta y \\
                 \delta y^* \\
               \end{array}\right),
\end{eqnarray}

where the $3\times3$ array $A$ is given by Eq. (\ref{A33}) and the
$2\times2$ array $B$ is given by Eq.(\ref{B22}), (see Appendix
\ref{sec:fluct})

\begin{figure*}[tb]
\begin{eqnarray}\label{A33}
A=\left(
  \begin{array}{ccc}
  (-2k_1-[k_2+k_3]\bar \chi
-\frac{a}{V}-\frac{a}{V^*}) &
(-(k_2+k_3)\bar A+(k_{-2}+k_{-3})\bar \chi+k_{-1}-\frac{a}{V^*}) & -\frac{a}{V} \\
   &   &   \\
(2k_1+[k_2+k_3]\bar \chi) & ((k_2+k_3)\bar
A-k_{-1}-(k_{-2}+k_{-3})\bar \chi-\frac{a}{V}) &
\frac{a}{V} \\
   &   &    \\
-\frac{V}{V^*}(2k^*_1+[k_2^*+k_3^*]\bar \chi^*) &
(\frac{a}{V^*}-\frac{V}{V^*}(2k^*_1+[k_2^*+k_3^*]\bar \chi^*) &
([k_2^*+k_3^*](\frac{C}{V^*}-\frac{V}{V^*}(\bar A+\bar \chi)-\bar \chi^*) \\
 &  & -k^*_{-1}-(k^*_{-2}+k^*_{-3})\bar \chi^* -\frac{a}{V^*}  \\
 &  & -(2k^*_1 + [k_2^*+k_3^*]\bar \chi^*))\\
  \end{array}
\right) \nonumber,
\\
\end{eqnarray}
\end{figure*}

\begin{figure*}[tb]
\begin{equation}\label{B22}
B=\left(
  \begin{array}{cc}
    \Big(-k_{-1}-k_{-2}\bar \chi +
(k_2-k_3)\bar A-\frac{a}{V}\Big) & \frac{a}{V} \\
    \frac{a}{V^*} & \Big(
(k_2^*-k_3^*)(\frac{C}{V^*}-\frac{V}{V^*}(\bar A+\bar \chi)-\bar
\chi^*)
-k^*_{-1}-k^*_{-2}\bar \chi^* -\frac{a}{V^*} \Big)\\
  \end{array}
\right).
\end{equation}
\end{figure*}

The Jacobian matrix $J$ (\ref{Jacobian1}) must be evaluated on
non-negative stationary solutions $\bar A \geq 0,\bar \chi \geq 0,
\bar \chi^* \geq 0$ corresponding to $\bar y = \bar y^*=0$. Then, to
assess the stability of the solution, the five eigenvalues
$\lambda_i, i= 1,2,...,5$ of the Jacobian must be calculated. If any
one of these five eigenvalues is positive (or their real part, if
complex) then the solution is unstable. This means that the system
will evolve to a chiral state, so mirror symmetry will be broken.
Only if \textit{all} five of the eigenvalues are negative (real
part) can we claim that the solution is stable. Deriving manageable
and useful closed form expressions for the eigenvalues $\lambda_i$
of Eqs. (\ref{A33},\ref{B22}) is practically impossible, due to the
fact that the racemic fixed point solutions $\bar A,\bar \chi, \bar
\chi^*$ lead to unwieldy expressions (as solutions of coupled
quartic equations). On the other hand, direct numeric calculation of
the fixed point solutions and their associated eigenvalues is
amenable and provides a wealth of information about the dynamic
stability of the underlying model, as functions of the chemical
rates and the system parameters. We will therefore map out regions
of stability/instability in parameter space. We carry this out
assuming immobilized catalyst from the start, setting $k^*_2,
k^*_{-2}$ to zero in one compartment, and $k_3, k_{-3}$ in the
other. Variations in the remaining rate constants are carried out
satisfying the constraints in Eqs. (\ref{inverse},\ref{microrev*}).

We apply a second, independent, stability test which does not
require calculating the eigenvalues: namely, the Routh-Hurwitz (RH)
criteria. We derive explicit expressions whose algebraic signs
indicate whether the racemic fixed point is stable or unstable. The
canonical form of the characteristic polynomial for the complete
$5\times 5$ jacobian Eq. (\ref{Jacobian1}) is
\begin{equation}\label{canonical}
P(\lambda) = \lambda^5 + a_1\lambda^4 + a_2\lambda^3 + a_3\lambda^2
+ a_4\lambda + a_5 = 0.
\end{equation}
Then (see Appendix B of \cite{Murray}) there are conditions on the
coefficients $a_i$, $i=1,2,...,5$ such that the zeros of
$P(\lambda)$ have $\Re \lambda < 0$. The necessary and sufficient
conditions for this to hold are the Routh-Hurwitz conditions. One
such form, together with
\begin{equation}\label{RHconditions5}
a_5 \equiv -\det\big(B\big)\det\big( A \big)> 0,
\end{equation}
is that
\begin{eqnarray}\label{RHconditions1}
D_1 &=& a_1 \equiv -[{\rm tr}\big(B\big) + {\rm tr}\big(A\big)]
> 0, \, \& \\\label{RHconditions2} D_2 &=& (a_1a_2-a_3)
> 0, \, \&
\\\label{RHconditions3}
D_3 &=& a_3D_2 + a_1(a_5 - a_1a_4) > 0, \, \&
\\\label{RHconditions4} D_4 &=& a_4 D_3 - a_5\{ a_1(a_2^2-a_4)
-(a_3a_2 - a_5)\} > 0.
\end{eqnarray}
The expressions $a_i$ can be read off directly from comparing the
polynomial in Eq.(\ref{canonical}) to $-P$ calculated in Eq.
(\ref{charpolyfull}).

If \textit{any} of the above conditions Eqs.
(\ref{RHconditions5}-\ref{RHconditions4}) does not hold, then the
racemic fixed point solution is unstable. As before, this means the
system will evolve to a chiral state. This test can be compared with
direct numerical calculation of the five eigenvalues $\lambda_i$
$i=1,2,3,4,5$ (the roots of the characteristic polynomial). We find
complete agreement between the two methods (eigenvalues, RH
criteria) employed. We emphasize that an instability in the racemic
fixed point implies the onset of a bifurcation to a non-racemic,
hence chiral, solution.

%------------------------------------------------------------------------
\subsection{\label{sec:graph} Domains of instability}
%------------------------------------------------------------------------

We initiate the procedure outlined above by specifying the
forward/reverse reaction rates (temperature differences are treated
implicitly), the internal flow rate, the compartment volumes and the
conserved total system mass:
\begin{equation}\label{params}
\{k_{\pm i}(T), k^*_{\pm i}(T^*)\}, a, V, V^*, C,
\end{equation}
the individual rates of course satisfying microreversibility at the
respective temperatures $T,T^*$. We then solve for the complete
racemic fixed point solution, retaining only those solutions that
are non-negative:
\begin{equation}
\left\{ \frac{dA}{dt} = 0, \frac{d \chi}{dt} = 0, \frac{d
\chi^*}{dt} = 0\right\}_{\bar y=\bar y^* = 0} \Rightarrow \{\bar
A,\bar \chi,\bar \chi^*\}_{\geq 0}.
\end{equation}
We next evaluate the Jacobian matrix over this fixed point solution:
\begin{equation}\label{Jacobian}
J \Rightarrow \left(
                \begin{array}{cc}
                  A^{3\times3} & 0 \\
                  0 & B^{2\times2} \\
                \end{array}
              \right)|_{ \{\bar A,\bar \chi,\bar \chi^* \}_{\geq 0} }.
\end{equation}
As a final step we evaluate the five RH conditions
Eqs.(\ref{RHconditions5}- \ref{RHconditions4}): are they
\textit{all} true or not? If not, then we immediately know that the
racemic fixed point is unstable for the parameter choice made in Eq.
(\ref{params}). Hence any fluctuation about the idealized racemic
composition will grow and drive the system to a chiral final state.
In parallel, we also evaluate numerically the five roots of the
characteristic polynomial $\lambda_i$ $i=1,2,3,4,5$ and verify
agreement between RH criteria and the eigenvalues. We use the RH
criteria to map out the regions of linear instability for the
racemic solution.

We start with the reaction rates $k_1 = 10^{-9}$, $k_{-1} =
10^{-15}$, $k_1^* = 10^{-7}$, $k_{-1}^* = 10^{-11}$, $k_2 =10^2$,
$k_{-2} = 10^{-4}$, $k_2^* = 0$, $k_{-2}^* = 0$, $k_3 = 0$, $k_{-3}
= 0$, $k_3^* = 10^1$, $k_{-3}^* = 10^{-3}$, the volume and flux
parameters $V = 10$, $V^* = 10$, $a = 0.1$, and the initial
concentrations $[A]_0 = 10^{-6}$, $[A^*]_0 = 10^{-6}$, $[L]_0 =
10^{-11}$, $[D]_0 = 10^{-11}$, $[L^*]_0 = 10^{-11} + 10^{-21}$,
$[D^*]_0 = 10^{-11}$ as employed in Ref \cite{vents}. This point in
parameter space was shown to lead to SMSB and subsequent chiral
amplification from direct numerical integration of the differential
rate equations. These specific rate values and system parameters
were obtained after performing a set of trial and error numerical
simulations obeying the condition $k^*_{-3}
> k_{-2}$ and $k_2 > k^*_3$ which corresponds
approximately to thermodynamically unattainable conditions for
systems with a uniform temperature and lacking compartments. Once
obtained, we can exploit the stability analysis to map out and
amplify the full domain of instability of the racemic solution.

\begin{figure*}[h]
\begin{center}$
\begin{array}{cc}
a) &
b) \\
 &
 \\
\includegraphics[height=3in]{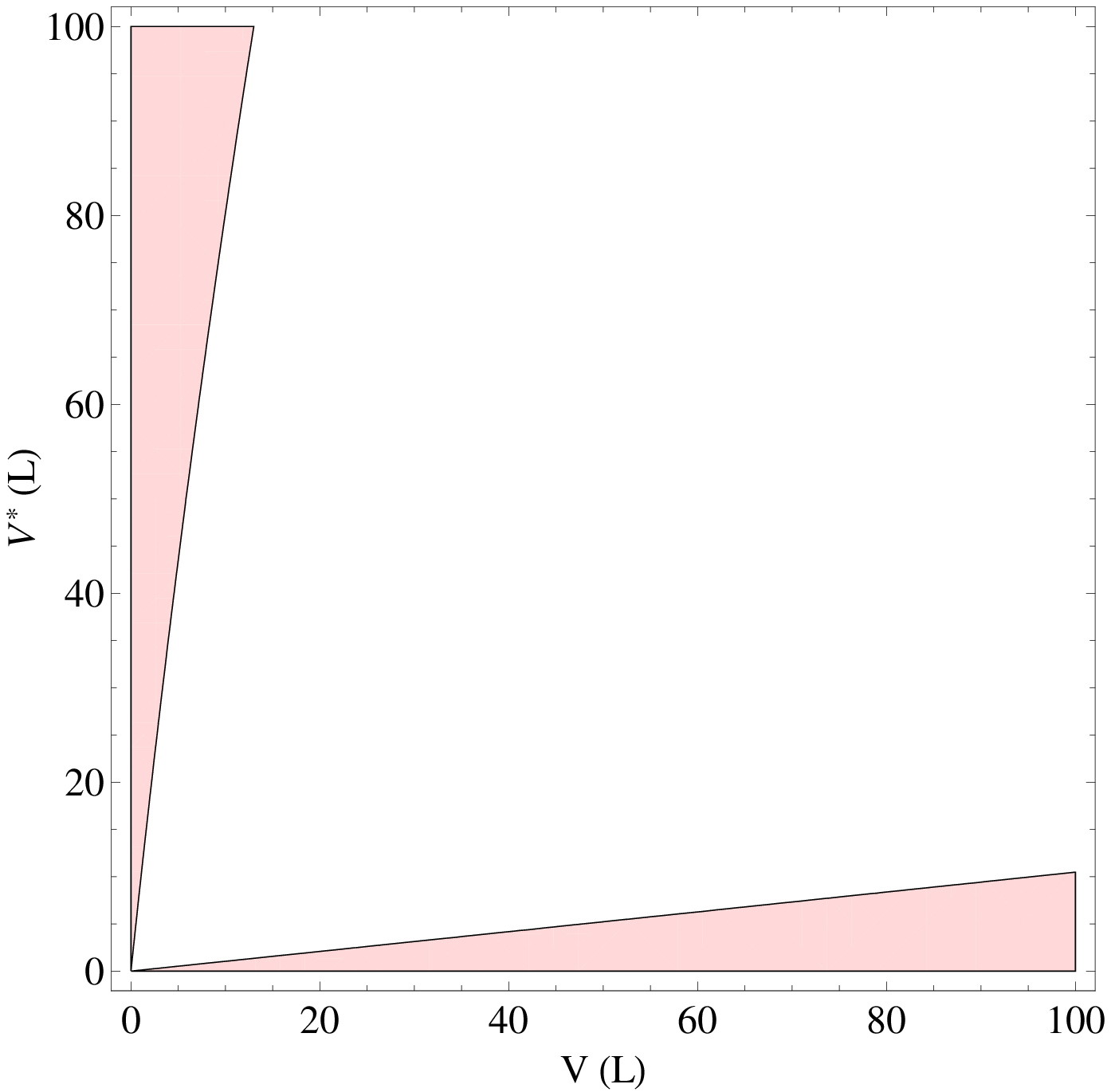} &
\includegraphics[height=3in]{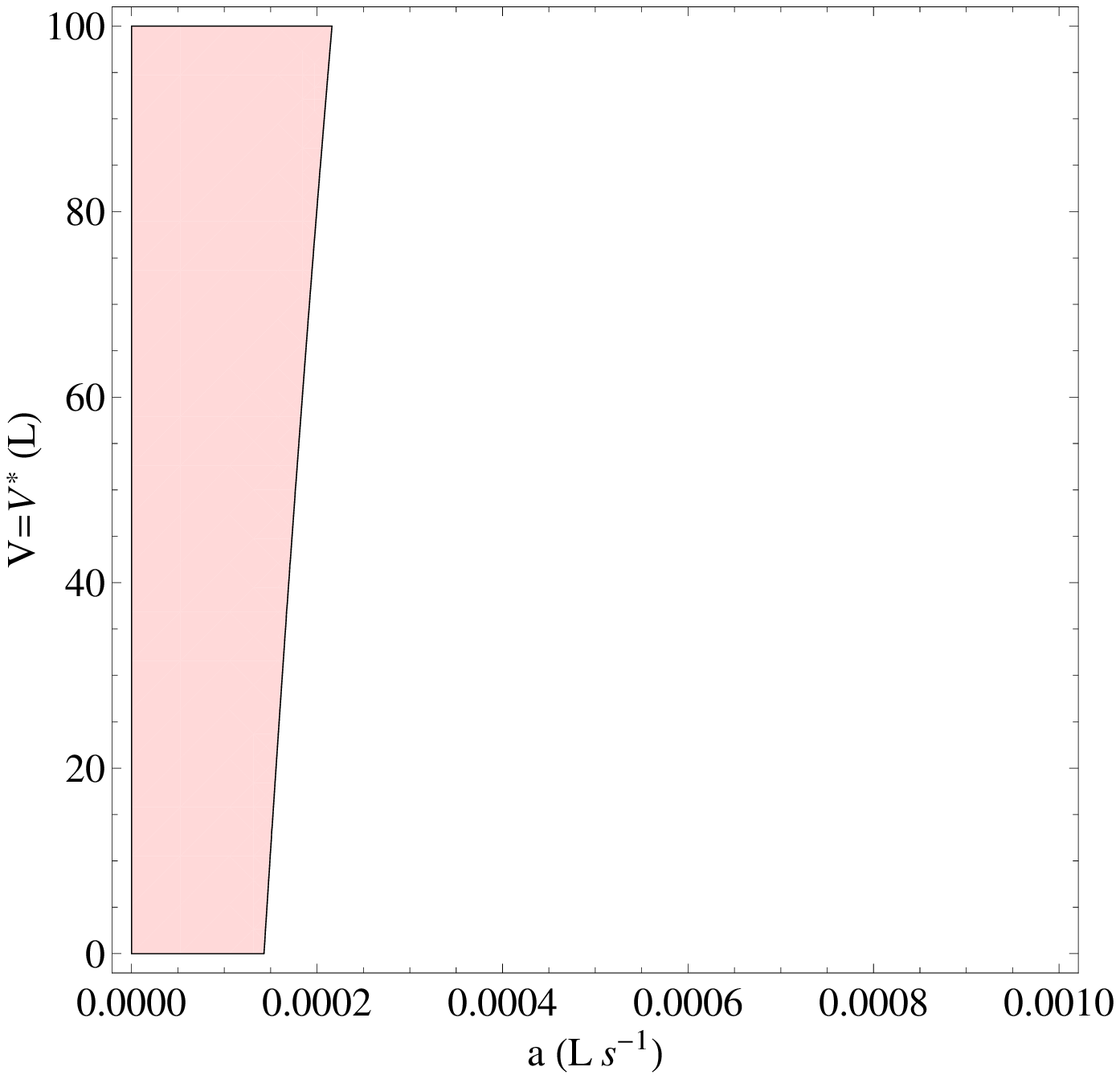} \\
 &
 \\
c) &
d) \\
 &
 \\
\includegraphics[height=3in]{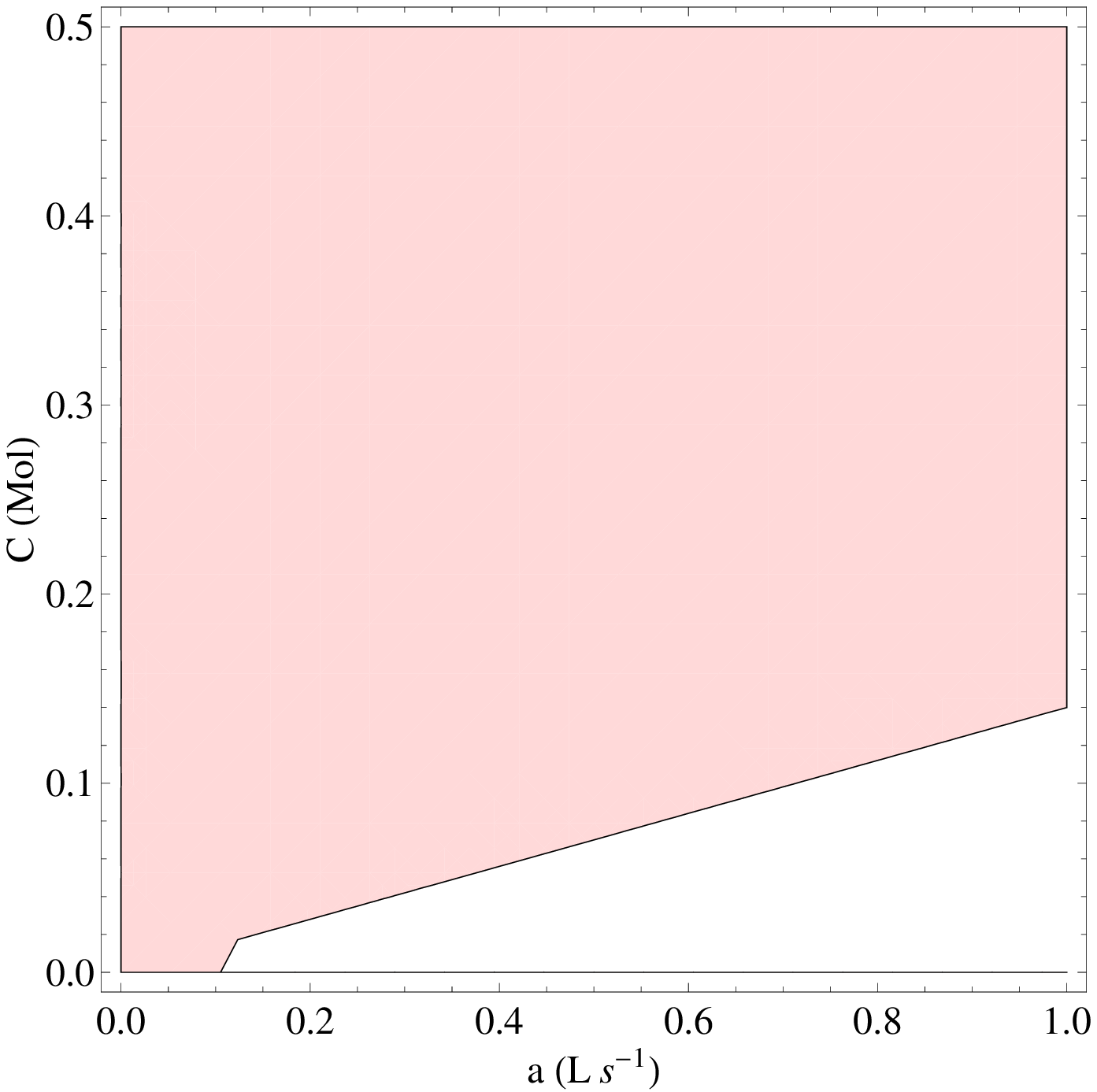} &
\includegraphics[height=3in]{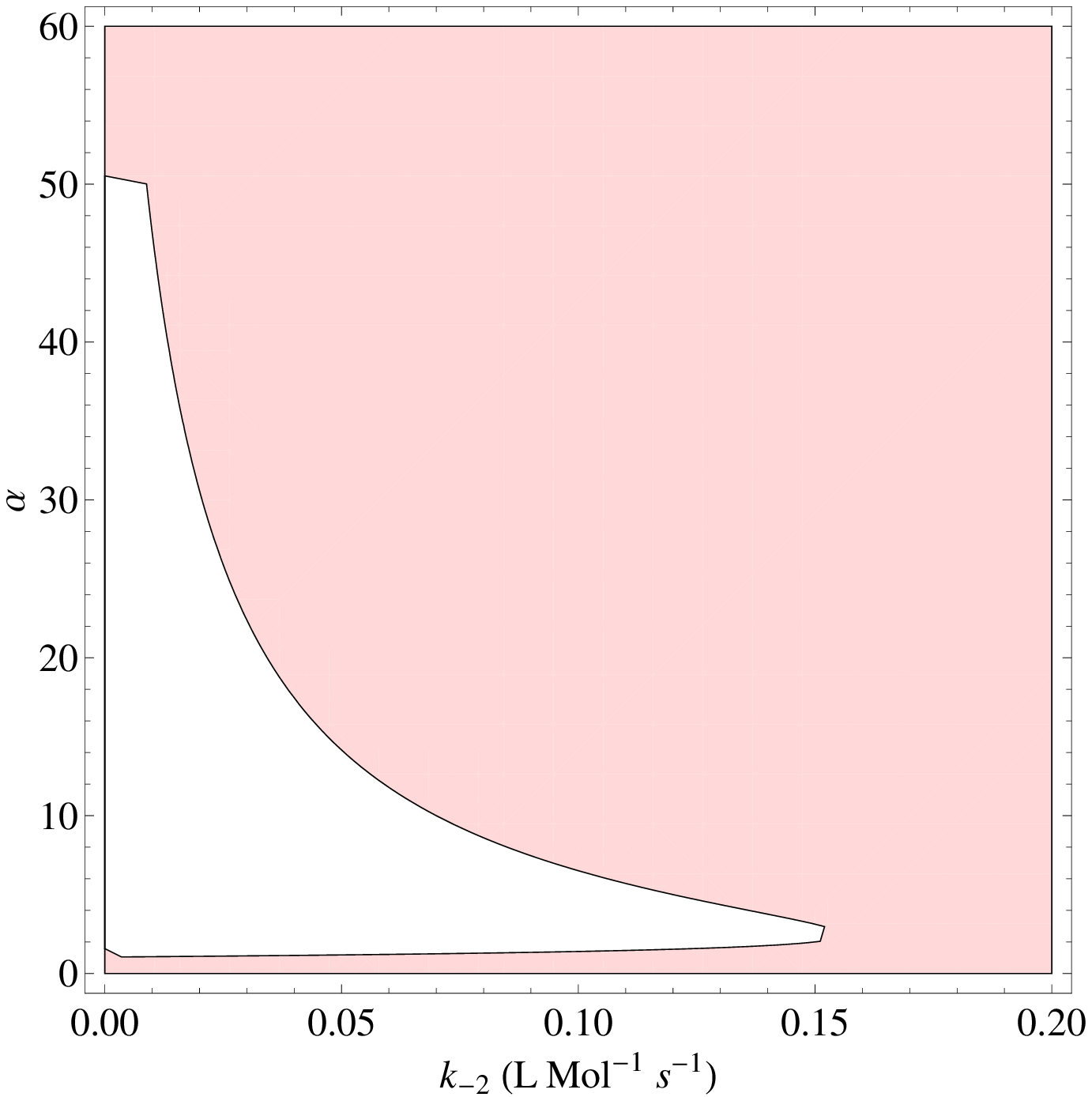}
\end{array}$
\end{center}
\caption{\label{firstset} White zones indicate where the racemic
state is \textbf{unstable} to perturbations and so bifurcates to a
chiral state (color/grayscale, where racemic state is stable) as a
function of the indicated selected pairs of variables: a) The cold
and hot compartment volumes $V$ and $V^*$, b) the flow parameter $a$
versus the volume of each compartment, holding $V=V^*$ fixed, c) the
flow parameter $a$ and the total system concentration $C$, and d)
the reaction rate $k_{-2}$ and $\alpha$, where
$\alpha=k_{-3}^*/k_{-2}$. Except for the specific pair that is
varied in a),b),c), and d), the remainder of values are held fixed
at $k_1 = 10^{-9}$, $k_{-1} = 10^{-15}$, $k_1^* = 10^{-7}$,
$k_{-1}^* = 10^{-11}$, $k_2 =10^2$, $k_{-2} = 10^{-4}$, $k_2^* = 0$,
$k_{-2}^* = 0$, $k_3 = 0$, $k_{-3} = 0$, $k_3^* = 10^1$, $k_{-3}^* =
10^{-3}$, the volume and flux parameters $V = 10$, $V^* = 10$, $a =
0.1$. The initial concentrations $[A]_0 = 10^{-6}$, $[A^*]_0 =
10^{-6}$, $[L]_0 = 10^{-11}$, $[D]_0 = 10^{-11}$, $[L^*]_0 =
10^{-11} + 10^{-21}$, $[D^*]_0 = 10^{-11}$.}
\end{figure*}

\begin{figure*}[h]
\begin{center}$
\begin{array}{cc}
\includegraphics[height=3in]{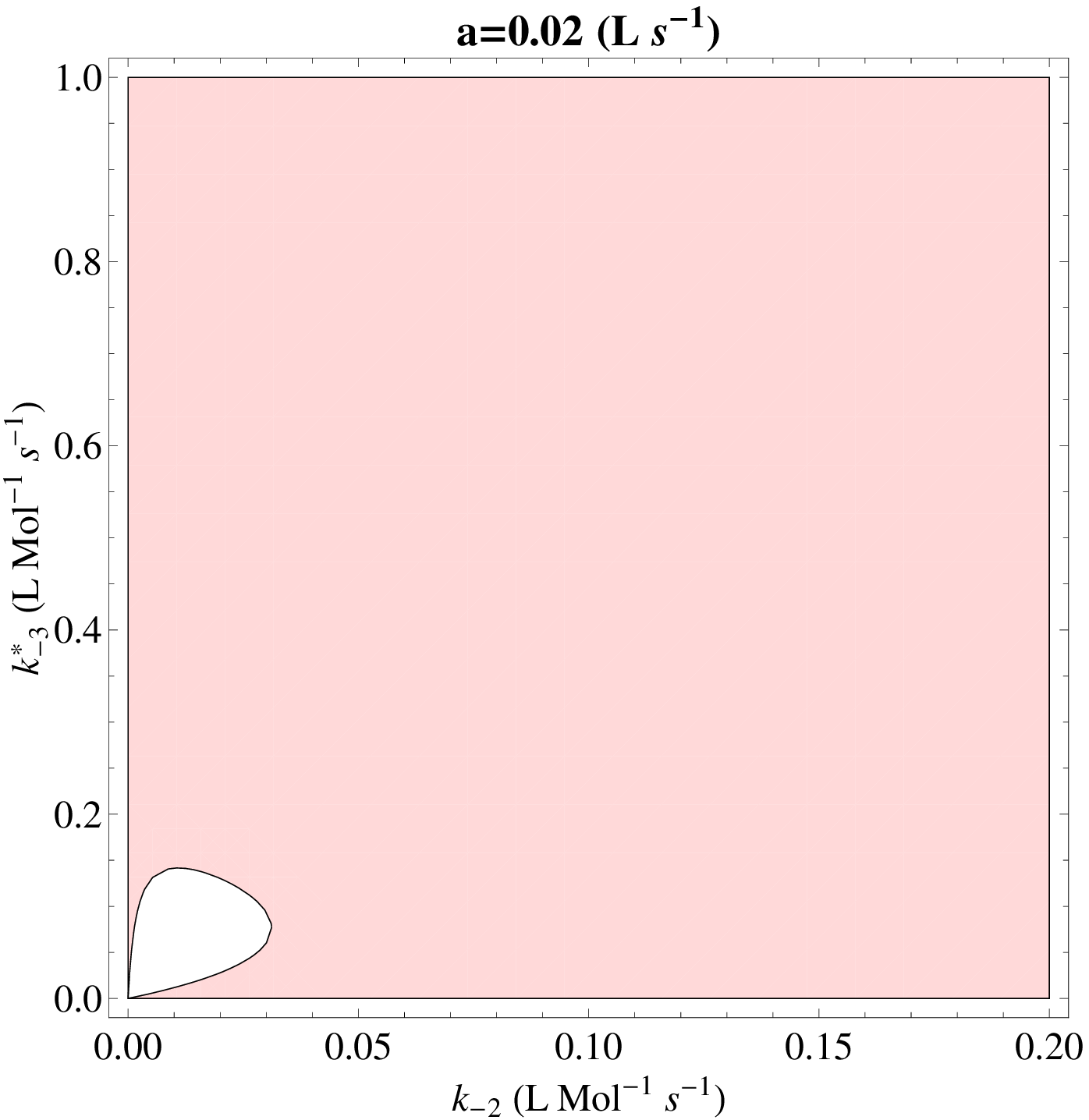} &
\includegraphics[height=3in]{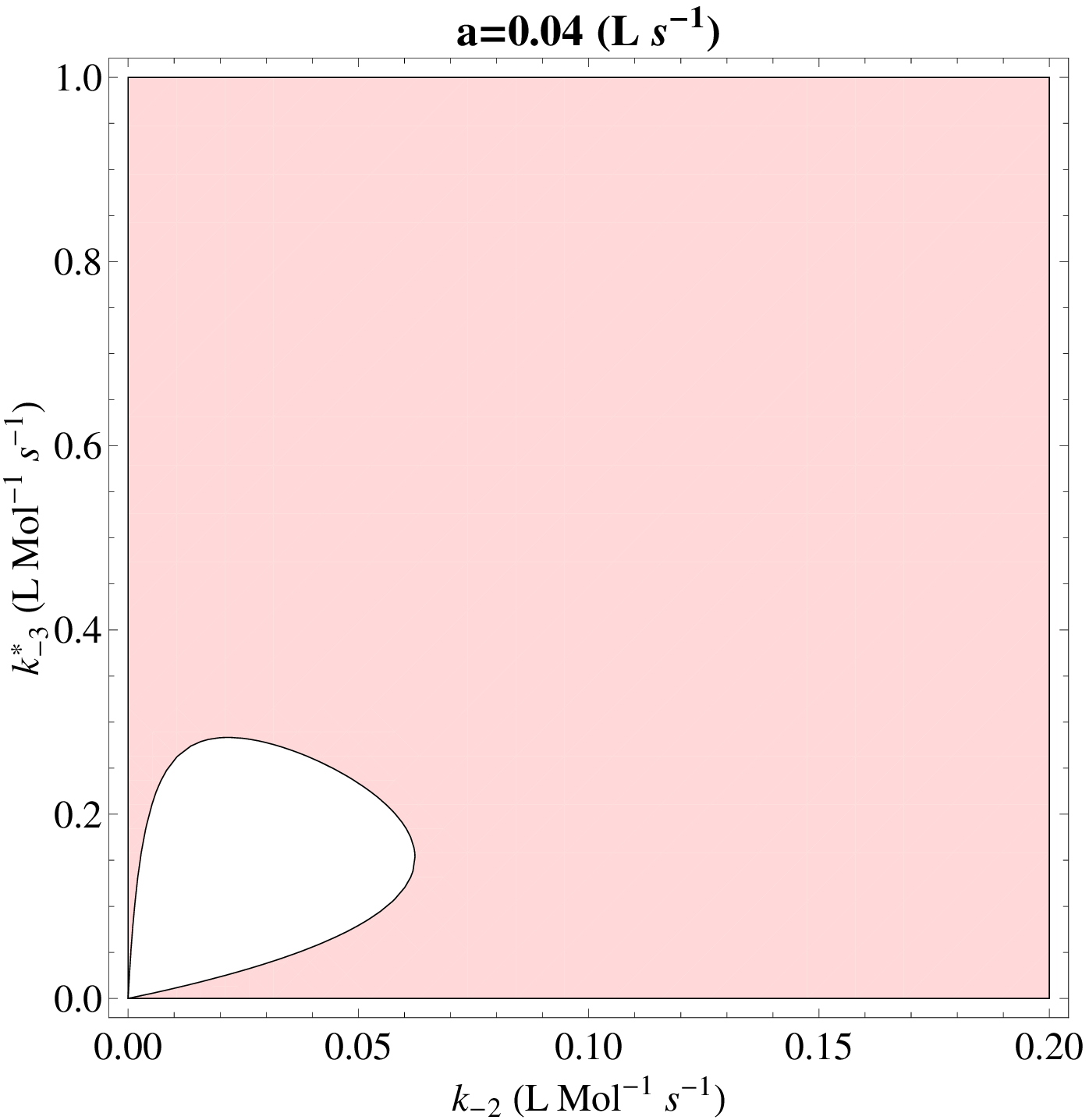} \\
 &
 \\
\includegraphics[height=3in]{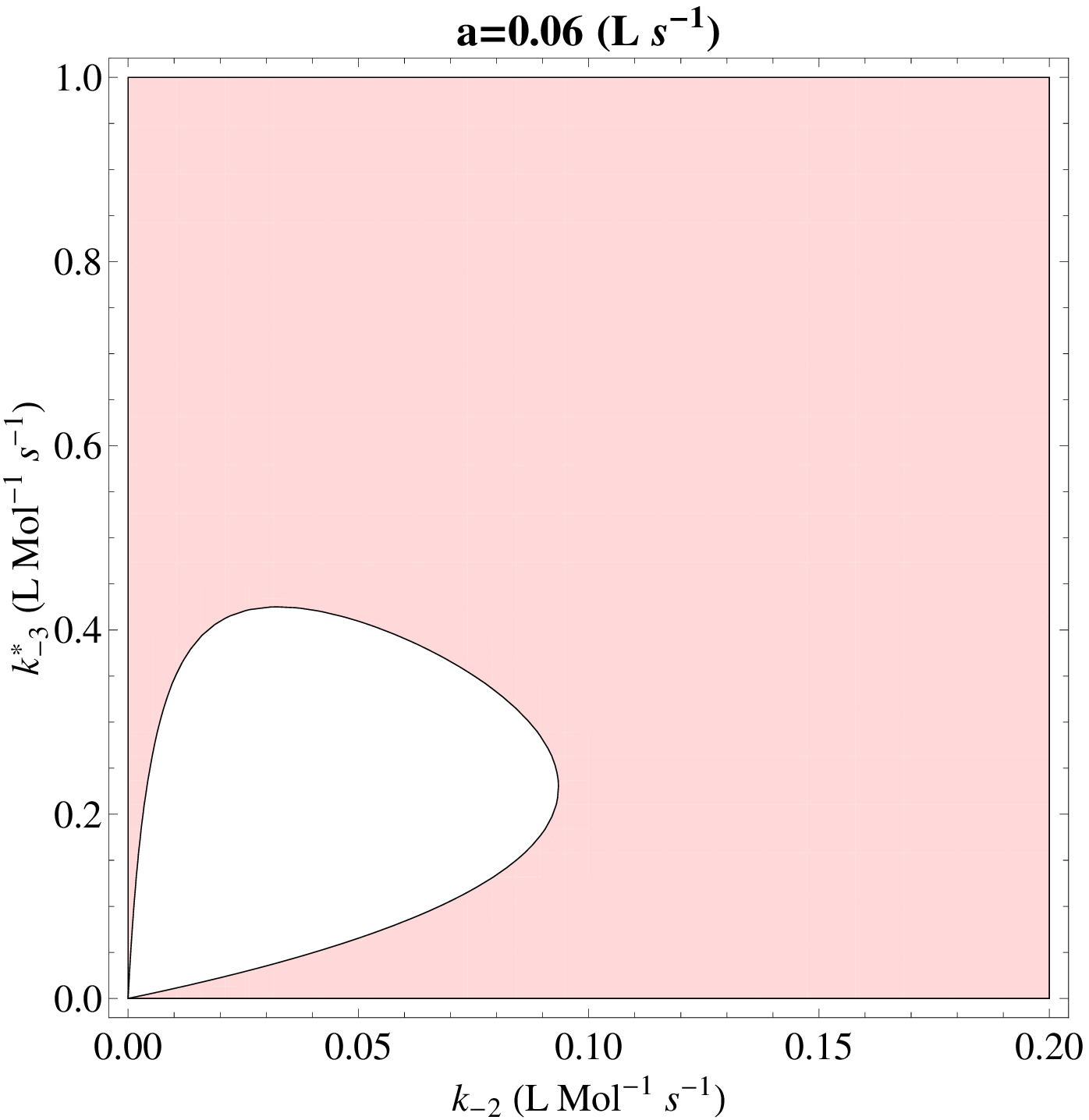} &
\includegraphics[height=3in]{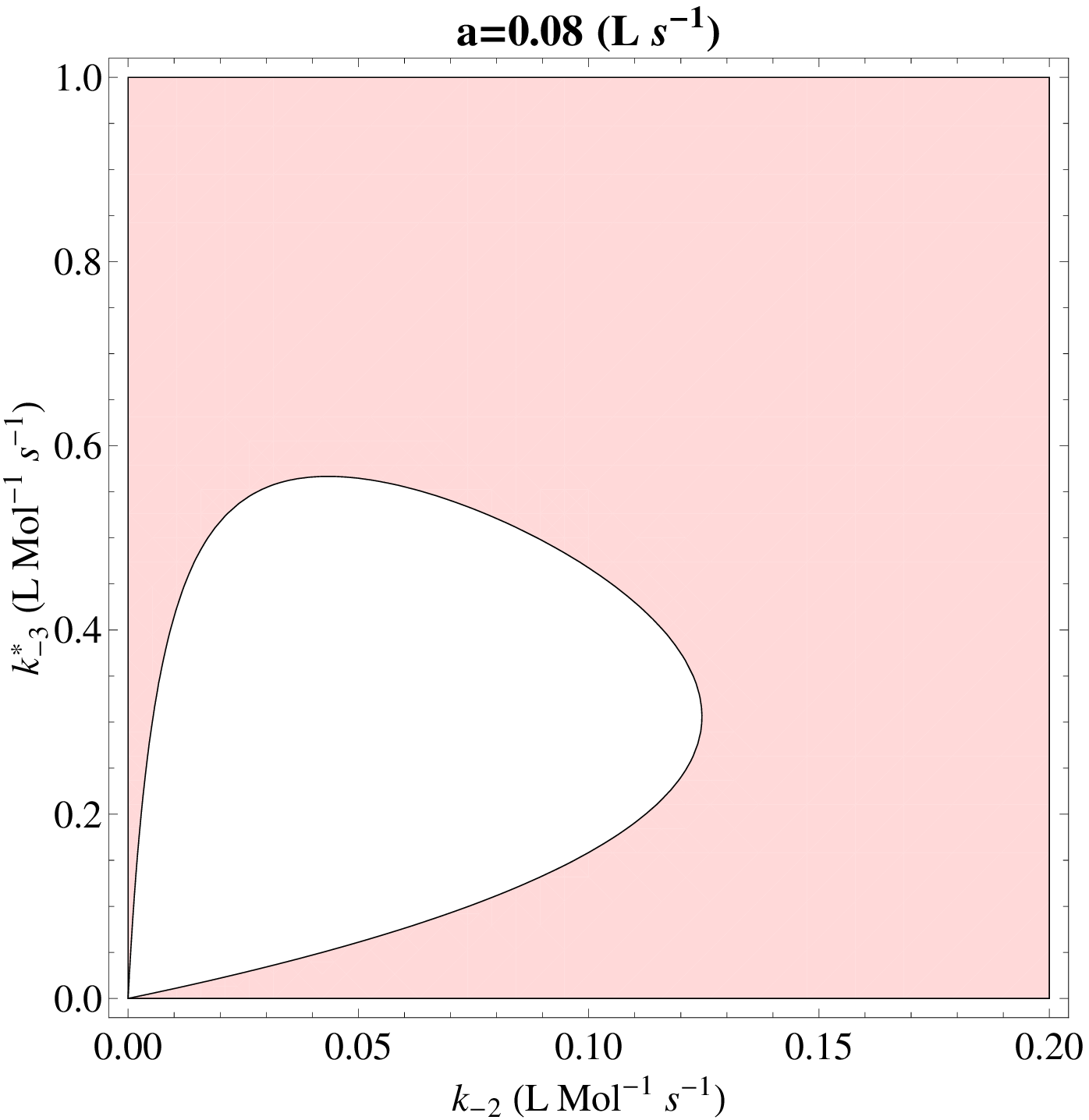} \\
 &
 \\
\end{array}$
\end{center}
\caption{\label{km2km3*b} White zones indicate where the racemic
state is \textbf{unstable} to perturbations (color/grayscale, where
its stable) for different values of $k_{-2}$ and $k_{-3}^*$, and for
different values of the internal flow parameter $a$. The area of the
region of instability scales up with the flow rate. The remainder of
parameters and initial concentrations as in Fig. \ref{firstset}.}
\end{figure*}

\begin{figure}[h]
\centering
\includegraphics[height=2.5in]{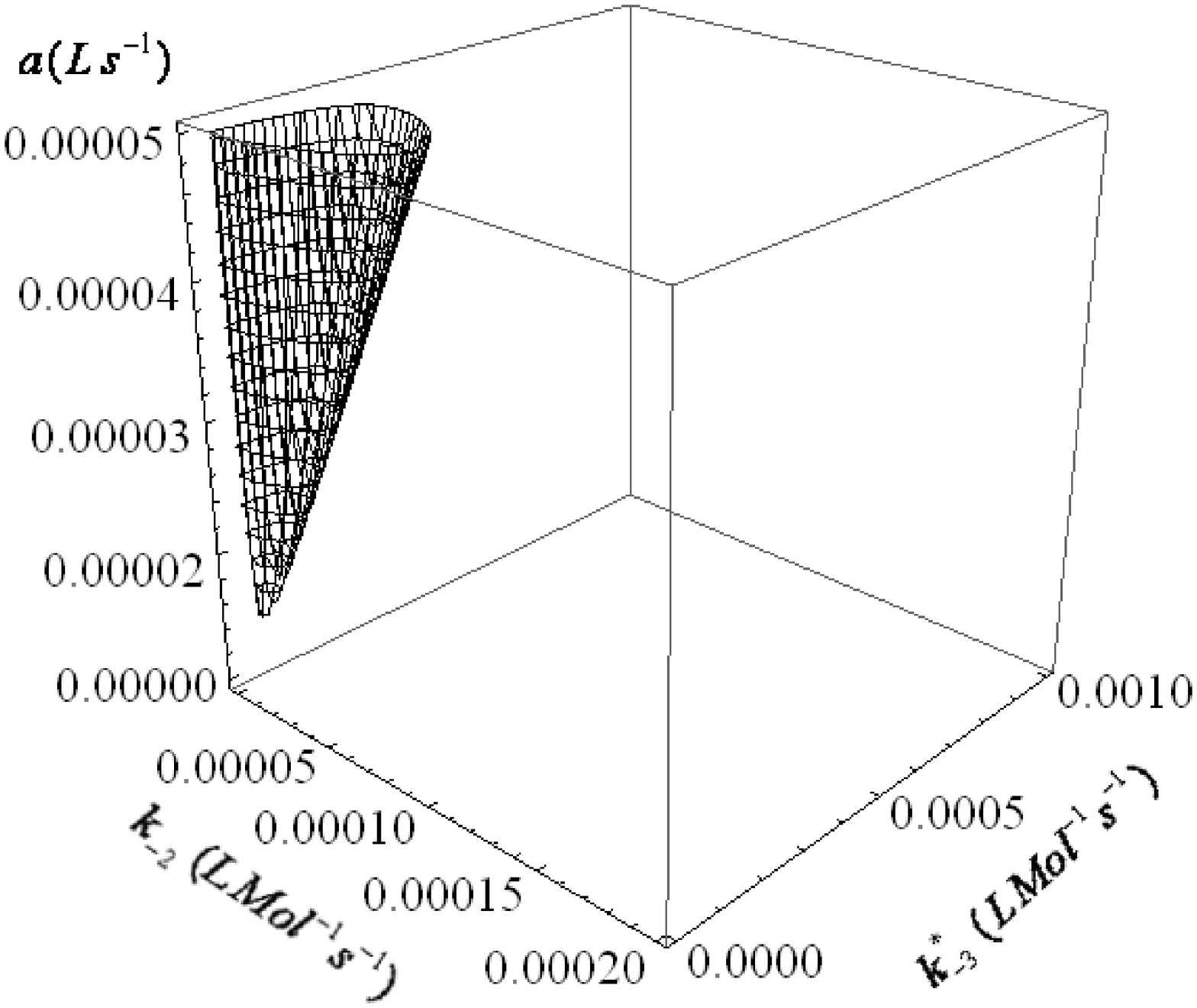}
\caption{\label{km2km3*3Da} Effect of varying the internal flow
rate. Three dimensional figure represents all points for which the
racemic state is \textbf{unstable} to perturbations. Increased flow
enlarges the allowed region of instability. The racemic state is
stable in the empty (white) domain. The remainder of parameters and
initial concentrations as in Fig. \ref{firstset}. }
\end{figure}

We consider how the stability of the racemic fixed point responds to
variations of selected pairs of variables about this point. In Fig
\ref{firstset} we display the regions of stability/instability as a
function of (a) the two compartment volumes $V$ and $V^*$, (b) the
flow rate $a$ versus equal compartment volumes $V=V^*$, (c) the
total concentration $C$ versus flow rate $a$, and (d) $\alpha =
\frac{k^*_{-3}}{k_{-2}}$ versus $k_{-2}$.  The graphs in (a) and (b)
merely tell us that the ratio of the compartment volumes cannot be
either arbitrarily large nor small, and that for equal compartment
volumes, there is a minimum flow rate below which no chiral state
can be obtained. Perhaps more surprising are the trends indicated in
Fig \ref{firstset} (c) and (d): for a given flow rate there is a
critical system concentration \textit{above} which mirror symmetry
cannot be broken and in (d), for a given value of $k_{-2}$ a bounded
region in $k^*_{-3}$ in which the racemic state is unstable, and
this region narrows down and pinches off for a critical value of
$k_{-2}$. We can here appreciate that the conjecture $k^*_{-3}
> k_{-2}$ is necessary, but certainly not sufficient to lead to an instability
in the racemic state.  We therefore delve further into this
nonlinear relationship by considering how the instability varies
with $k^*_{-3},k_{-2}$ and with the flow rate $a$. In Fig
\ref{km2km3*b} we display a sequence of plots showing how the region
of instability grows in area as we scale up the flow rate. For a
given flow rate, there is a maximum value of $k_{-2}$ beyond which
no value of $k^*_{-3}$ will lead to an instability, and this maximum
value scales with $a$. Below this maximum value, there are always
upper and lower bounds on $k^*_{-3}$ between which the system is
unstable. These bounds also scale with $a$. In Fig \ref{km2km3*b} we
keep both $k_{-3}^*$ and $k_{-2}$ fixed and vary $a$: increased flow
enlarges the region of instability. This can also be appreciated in
the three-dimensional plot in Fig \ref{km2km3*3Da} exhibiting how
the region of instability expands in cross sectional area as we
scale up the internal flow rate $a$.

\begin{figure*}[h]
\begin{center}$
\begin{array}{cc}
\includegraphics[height=3in]{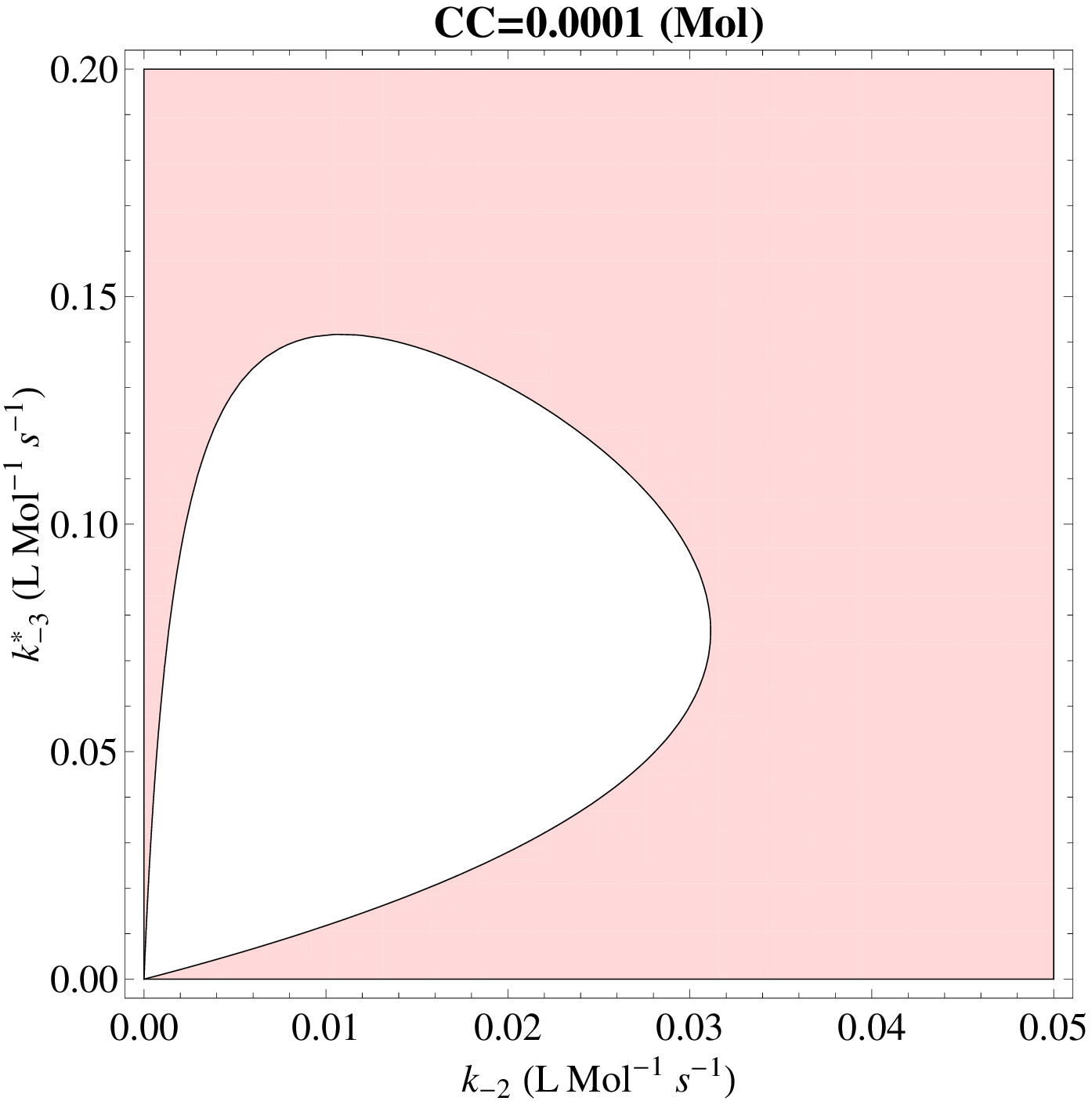} &
\includegraphics[height=3in]{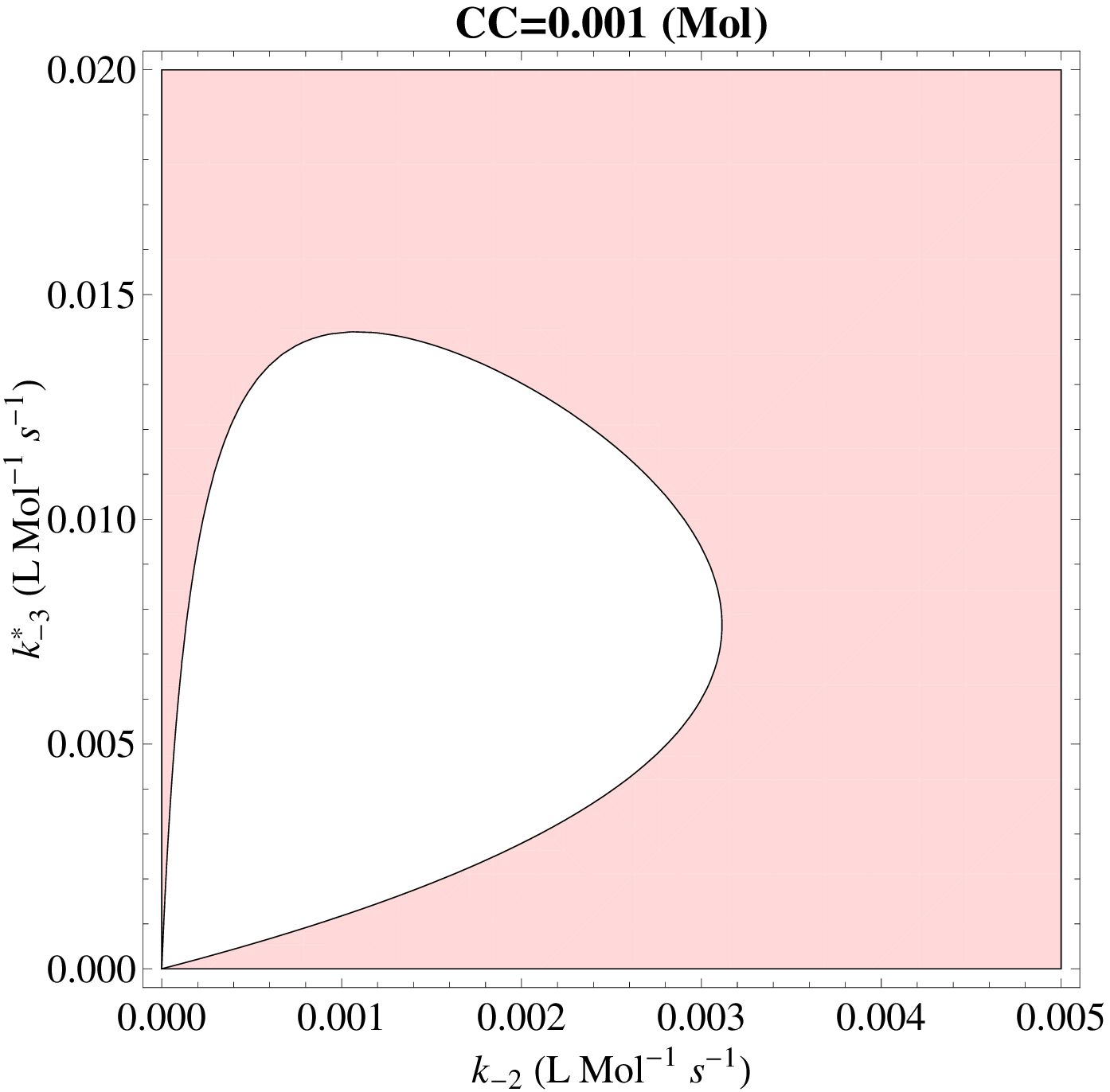} \\
 &
 \\
\includegraphics[height=3in]{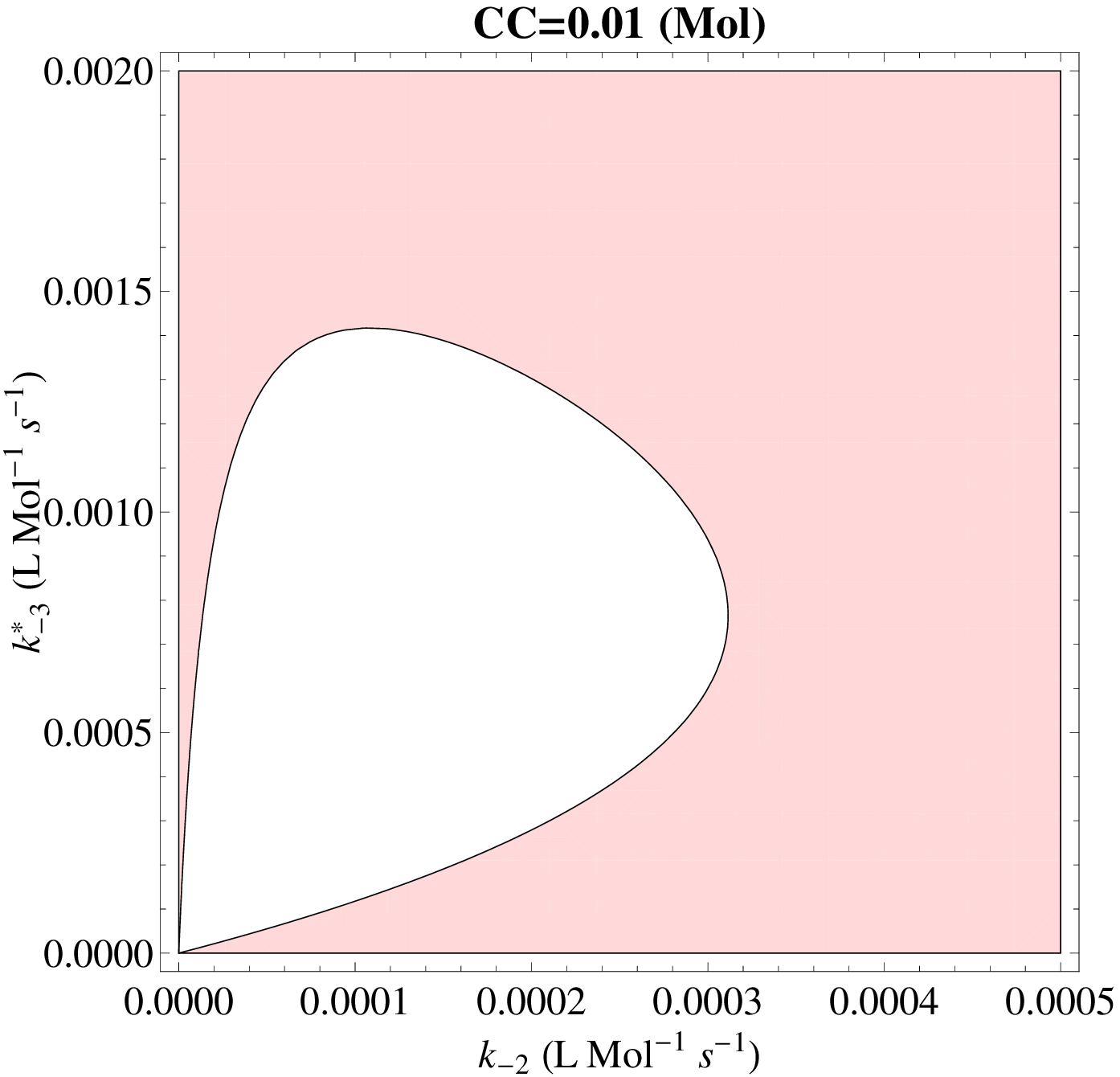} &
\includegraphics[height=3in]{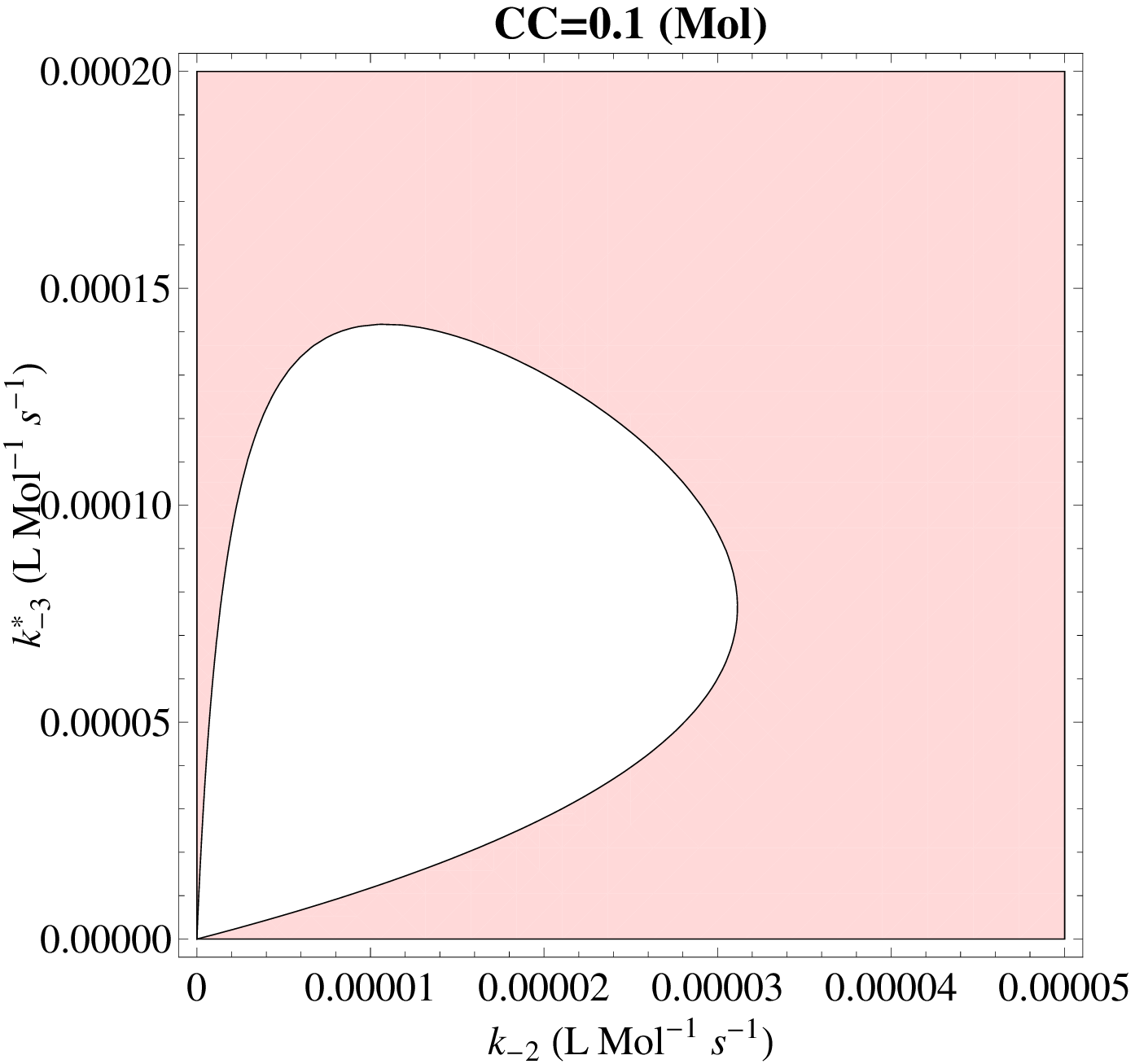} \\
 &
 \\
\end{array}$
\end{center}
\caption{\label{km2km3*d} Regions where the racemic state bifurcates
to chiral states (in white) for different values of $k_{-2}$ and
$k_{-3}^*$, and for different values of the total concentration $C$.
Increasing $C$ decreases the ranges in both $k_{-2}$ and $k_{-3}^*$
for which the racemic state is linearly unstable. The remainder of
parameters and initial concentrations as in Fig. \ref{firstset}.}
\end{figure*}

\begin{figure*}[h]
\begin{center}$
\begin{array}{cc}
\includegraphics[height=3in]{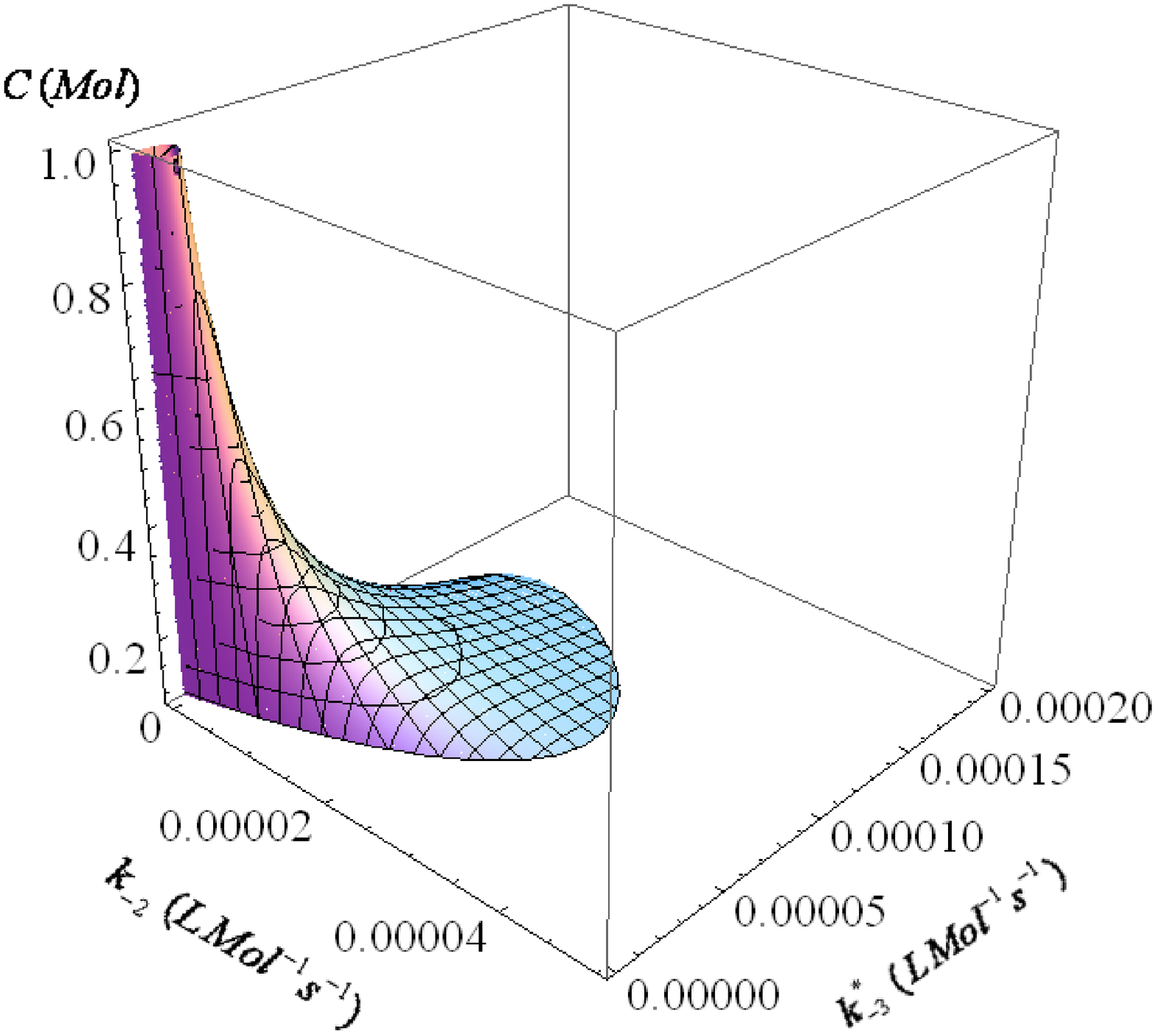}&
\includegraphics[height=3in]{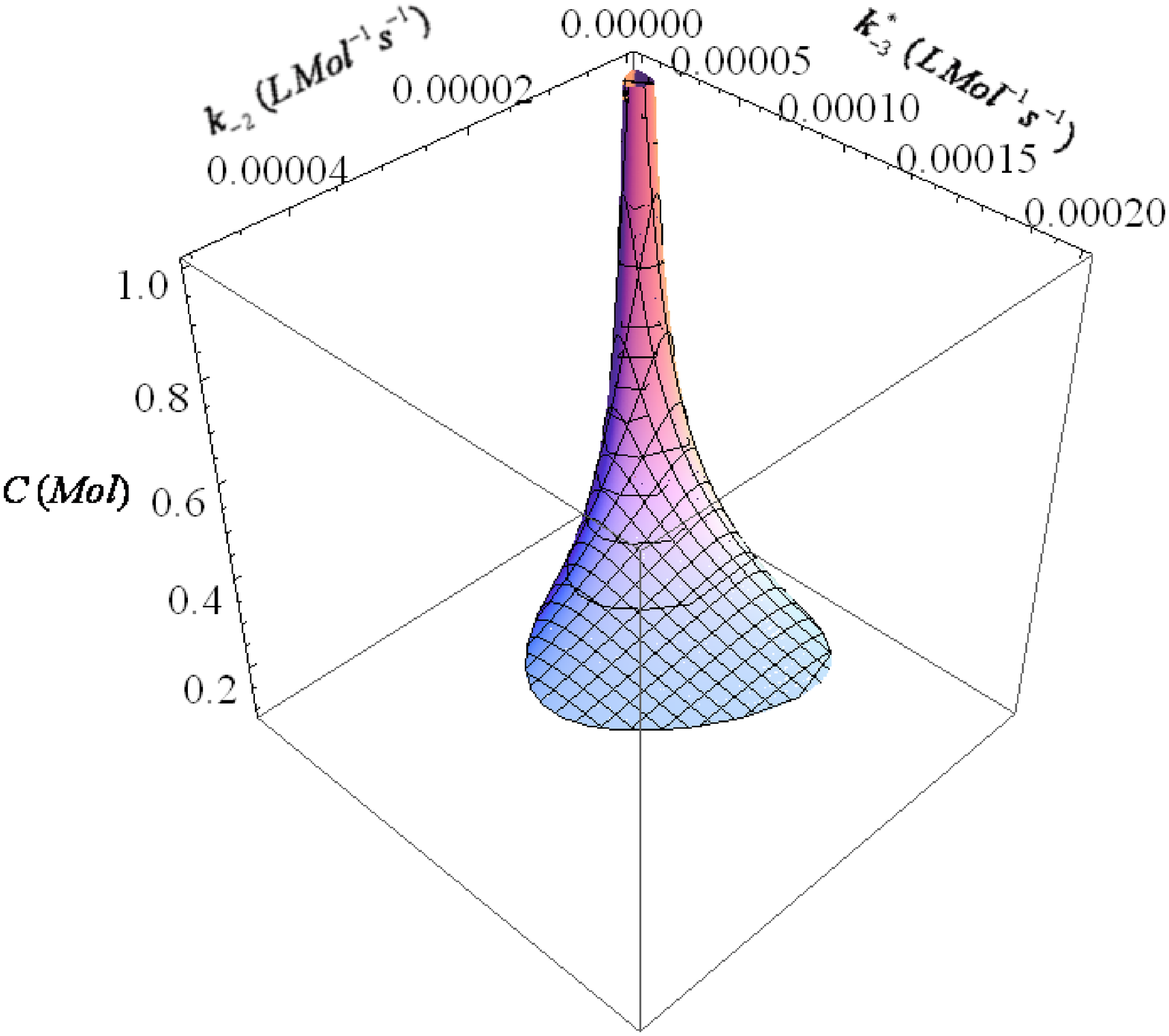} \\
\end{array}$
\end{center}
\caption{\label{km2km3*3Db} Effect of varying the total system
concentration $C$. The solid three-dimensional figure (color/gray)
indicates where the racemic state is \textbf{unstable} to
perturbations. Increasing the system concentration $C$
\textit{shrinks} the cross sectional domain $k_{-2}, k_{-3}^*$ of
instability. The racemic state is linearly stable in the empty
(white) region. The remainder of parameters and initial
concentrations are as listed in Fig. \ref{firstset}.}
\end{figure*}

In Fig \ref{km2km3*d} we display a sequence of plots showing how the
region of instability decreases in area as we scale up the total
concentration $C$, over three orders of magnitude. As $C$ is scaled
up, the rates $k^*_{-3},k_{-2}$ decrease in magnitude in such a way
as to preserve the \textit{shape} of the region of instability. As
before, for a given total concentration, there is a maximum value of
$k_{-2}$ beyond which no value of $k^*_{-3}$ will lead to an
instability, and this maximum value inversely scales with $C$. Below
this maximum value, there are always upper and lower bounds on
$k^*_{-3}$ between which the system is unstable. These bounds also
inversely scale with $C$. A three-dimensional plot in Fig
\ref{km2km3*3Db} indicates the cross-sectional area of the domain of
instability shrinks as we scale up the total system concentration
$C$. Dilute concentrations are more favorable for SMSB \cite{vents}.

%

%------------------------------------------------------------------------
\section{\label{sec:sims} Symmetry breaking and entropy production}
%------------------------------------------------------------------------

\begin{figure}[h]
\centering
\includegraphics[width=0.45\textwidth]{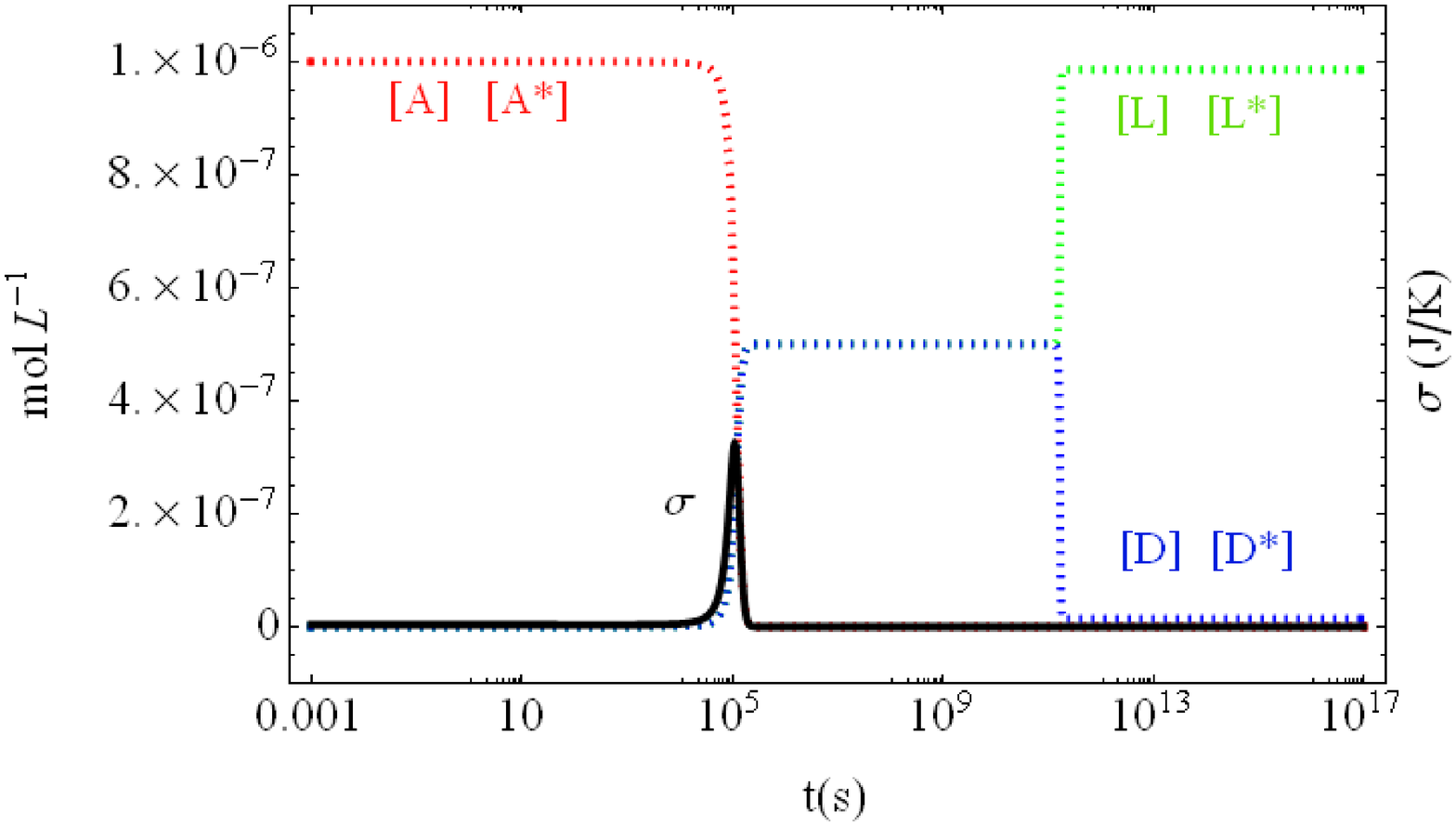}
\caption{\label{symbreak1} Symmetry breaking bifurcation in the two
compartment model, see Figure \ref{twocomps}. For the rates and
system parameters in given in Sec. \ref{sec:graph}.  The entropy
production $\sigma$ peaks at the onset of the induction period, well
before the symmetry breaking bifurcation. Note $\sigma
> 0$ for all times.}
\end{figure}

\begin{figure}[h]
\centering
\includegraphics[width=0.45\textwidth]{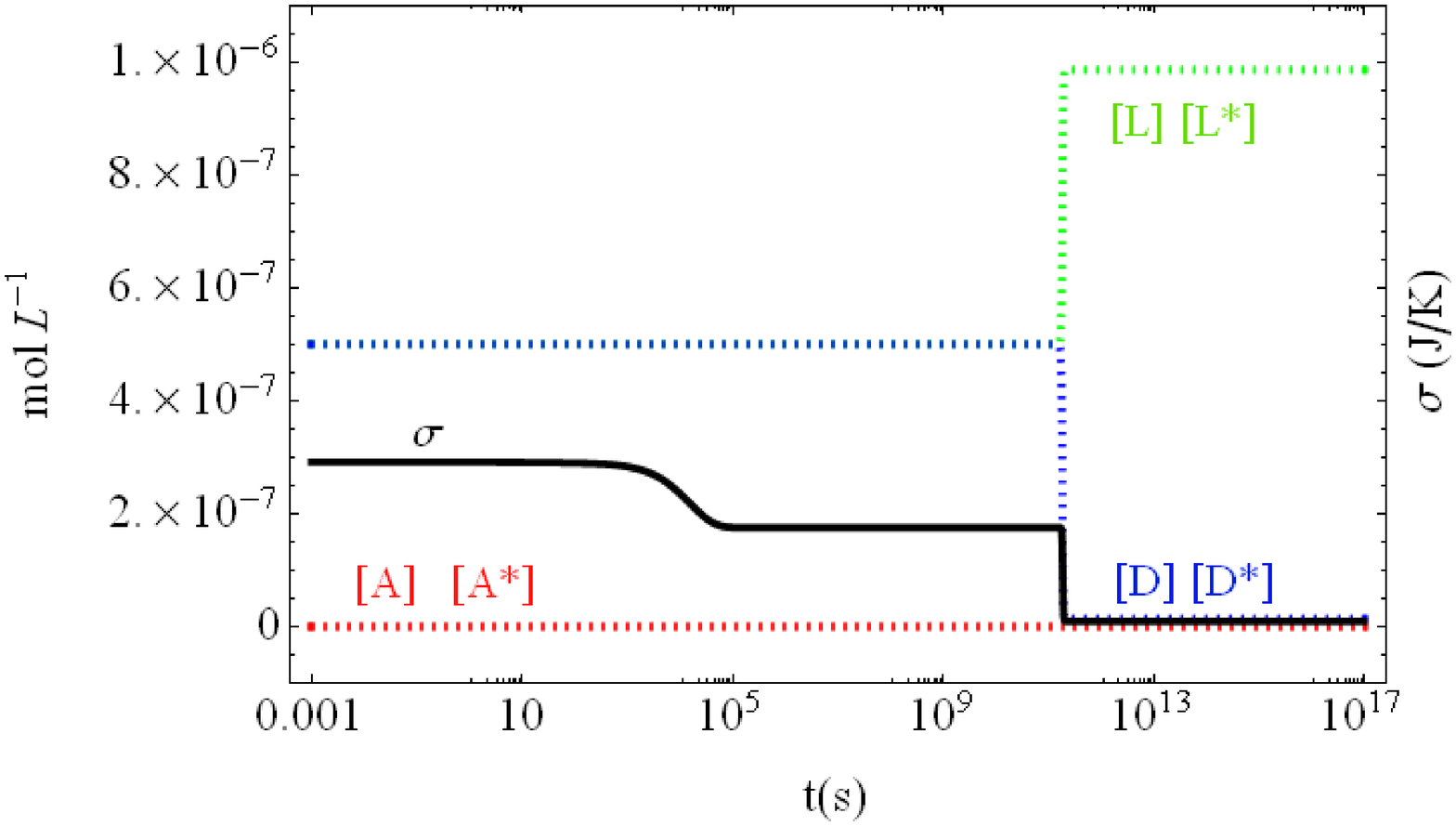}
\caption{\label{symbreak2} Symmetry breaking bifurcation in the two
compartment model, same rates and system parameters as in Figure
\ref{symbreak1}. See text for the initial concentrations. In this
case the entropy production $\sigma$ (scaled up by a factor of
$10^7$) decreases monotonically and drops to a minimum (but nonzero)
value at the symmetry breaking bifurcation, and remaining positive
for all subsequent times.}
\end{figure}

We illustrate the above considerations by way of two examples. The
inherent chiral fluctuations about the ideal racemic composition can
be modeled by starting with an initial ee below the statistical
deviation. In the first simulation, Figure \ref{symbreak1}, we begin
with $[A]_0 = [A^*]_0 = 1\times 10^{-6}M$, $[L]_0=[D]_0 =1 \times
10^{-11} M$, $[L^*]_0 = 1\times 10^{-6} + 1\times 10^{-21} M$, and
$[D^*]_0 = 1\times 10^{-6}M$. In the second, Figure \ref{symbreak2},
we keep the same rates and system parameters but start off with
different initial concentrations: $[A]_0 = [A^*]_0 = 1\times
10^{-11}M$, $[L]_0 = [D]_0 =5 \times 10^{-7} M$, $[L^*]_0 = 5\times
10^{-7} + 1\times 10^{-20} M$, and $[D^*]_0 = 5\times 10^{-7}M$.

We calculate and superimpose the entropy
production\cite{KondepudiPrigogine} $\sigma$ on the concentration
curves. The entropy production is a measure of the dissipation of
the out-of-equilibrium process under study. For the initial
conditions leading to Figure \ref{symbreak1}, the entropy production
suffers a peak at the onset of the induction period, reflecting the
fact that the catalysis of enantiomers Eq. (\ref{autoLD2}), and
subsequent depletion of achiral substrate, is the most important
contribution. The production then falls to a small but constant
nonzero value and remains nonzero as long as the system is kept out
of equilibrium. In the case of Figure \ref{symbreak2}, the entropy
production starts off from a large value then drops (at the same
time scale as the peak in Figure \ref{symbreak1}) and subsequently
drops down to its minimum nonzero value at the symmetry breaking
bifurcation. It again remains small but nonzero as long as the
system is kept out of equilibrium.

%------------------------------------------------------------------------
\section{\label{sec:disc} Concluding Remarks}
%------------------------------------------------------------------------

Compartmentalization, together with different temperatures for the
enantioselective and the non-enantioselective autocatalyses, was
shown to be a necessary condition for SMSB. The temperature gradient
and internal flow or recycling of hot and cold material is the
required driving force keeping the system far from equilibrium. On
the basis of the stability analysis of the racemic final state, the
results presented here demonstrate the existence of SMSB as a closed
region (see Figs. 3, 5 and 6) in the phase representation of the
reaction parameters. Explicit examples of SMSB  are displayed in
Figures \ref{symbreak1} and \ref{symbreak2} showing also the time
dependence of the associated production of entropy for starting from
two different initial concentrations.

The considerations given to the microreversibility condition in Secs
\ref{sec:intro}, \ref{sec:LES-Tgrad} and \ref{sec:arrhenius} serve
to underscore the fact that limited enantioselectivity-of and by
itself-- cannot lead to spontaneous mirror symmetry breaking,
neither for a uniform temperature (i.e. at an experimental condition
required to attain chemical equilibrium), not even for two
temperatures (i.e. an experimental condition that excludes chemical
equilibrium). By contrast, mirror symmetry breaking may be possible
when in addition to a non-uniform temperature distribution the
reactions (ii) and (iii) occur in two distinct compartments held at
different temperatures. The microreversibility arguments
\cite{RH,PigsFly}, used as a proof of the correctness of the model,
are necessary because we assume a scenario where chemical kinetics
can be applied.

We  briefly point out what modifications would have to be made in
order to consider spatial temperature gradients. First, we need to
input the temperature profile $T(x,y,z)$ as a known field. This
converts the reaction rates $k_i$ into spatially dependent functions
via the Arrhenius relation. The kinetic rate equations must be
replaced by partial differential reaction-diffusion-advection
equations for the spatial and temporal dependence of the
concentrations. The background convective flow of hot/cold material
could be modeled by a steady hydrodynamical flow-field compatible
with the compartment boundary conditions. In short, the technical
complications would be considerable, but not insurmountable.

The significance of the above results with respect to applied
absolute synthesis is somewhat limited due to the temperature
difference required for achieving the essential inequality $k^*_{-3}
> k_{-2}$. The point is, a sufficiently large difference
would require high pressure conditions as well to maintain the media
in a liquid state. By contrast, the results can be of significance
in scenarios of prebiotic chemistry. In this context, the required
experimental conditions of high temperature gradients and
compartmentalization agree with those found in deep ocean
hydrothermal vents, as we have previously reported \cite{vents}.
Furthermore, the more relevant feature of this LES scenario is its
ability to exhibit SMSB at very low reactant concentrations. In this
respect, the detailed description given here for the effect on the
SMSB dependence of the minimal internal flow rate -with respect to
the compartment volumes at different temperatures- and on the
reaction rate inequality $k^*_{-3}
> k_{-2}$, shows that the variability range for SMSB of these parameters
increases when the total system concentration $C$ decreases, i.e.
the probabilities for SMSB to occur \textit{increase} in a prebiotic
scenario. Notice that this overcomes the more important difficulty
for a reasonable SMSB in a prebiotic scenario, where a specific
prebiotic organic compound could be present, but only at very low
concentrations. This, in spite of the presence of an important
fraction of organic compounds, because of the diversity of different
organic compounds in prebiotic scenarios \cite{SchmittK}. While LES
is ruled out as a scheme for SMSB in experimental conditions where
thermodynamic equilibrium can be achieved, the results presented
here open up the study for SMSB in scenarios of non-uniform
temperature distributions, specific energy inputs to some species of
the system, and compartmentalization, i.e. in conditions similar to
those of living systems.

\section*{Acknowledgements}
The research of CB and DH is supported in part by the Grant
AYA2009-13920-C02-01, and that of JMR, JC, ZE-H and AM by
AYA2009-13920-C02-02, from MICINN (currently MINECO).

%-------------------------------------------------------------------
\appendix
\section{\label{sec:fluct} Fluctuation equations}
%------------------------------------------------------------------
The linearized fluctuation equations that follow from the kinetic
equations Eqs. (\ref{dAd}-\ref{detal}) in Sec.
\ref{sec:newvariables} are as follows. The overbar denotes a
stationary solution of Eqs. (\ref{dAd}-\ref{detal}).
\begin{eqnarray}
\bm{\dot{\delta A}} &=& \big(-2k_1-[k_2+k_3]\bar \chi
-\frac{a}{V}-\frac{a}{V^*}\big)\bm{\delta A} + \Big(-(k_2+k_3)\bar
A+(k_{-2}+k_{-3})\bar \chi+k_{-1}-\frac{a}{V^*}\Big)\bm{\delta
\chi} \nonumber \\
&+& (k_{-2}-k_{-3})\bar y \,\bm{\delta y} - \frac{a}{V}\bm{\delta
\chi^*}
\end{eqnarray}

\begin{eqnarray}
\bm{\dot{\delta \chi}} &=& (2k_1+[k_2+k_3]\bar \chi)\bm{\delta A}+
\Big((k_2+k_3)\bar A-k_{-1}-(k_{-2}+k_{-3})\bar
\chi-\frac{a}{V}\Big)\bm{\delta \chi}+ (k_{-3}-k_{-2})\bar y \,
\bm{\delta y} + \frac{a}{V}\bm{\delta \chi^*}
\end{eqnarray}

\begin{eqnarray}
\bm{\dot{\delta y}} &=& (k_2-k_3)\bar y \bm{\delta A} -k_{-2}\bar y
\bm{\delta \chi} +\Big(-k_{-1}-k_{-2}\bar \chi + (k_2-k_3)\bar
A-\frac{a}{V}\Big)\bm{\delta y}+ \frac{a}{V}\bm{\delta y^*}
\end{eqnarray}

\begin{eqnarray}
\bm{\dot{\delta \chi^*}} &=& -\frac{V}{V^*}(2k^*_1+[k_2^*+k_3^*]\bar
\chi^*)\bm{\delta
A}+\Big(\frac{a}{V^*}-\frac{V}{V^*}(2k^*_1+[k_2^*+k_3^*]\bar
\chi^*\Big)\bm{\delta
\chi}\nonumber \\
&+&\Big([k_2^*+k_3^*](\frac{C}{V^*}-\frac{V}{V^*}(\bar A+\bar
\chi)-\bar \chi^*)-k^*_{-1}-(k^*_{-2}+k^*_{-3})\bar \chi^* -
\frac{a}{V^*} -(2k^*_1 + [k_2^*+k_3^*]\bar \chi^*)\Big)\bm{\delta
\chi^*} \nonumber\\
&+& (k^*_{-3}-k^*_{-2})\bar y^* \,\bm{\delta y^*}
\end{eqnarray}

\begin{eqnarray}
\bm{\dot{\delta y^*}} &=& -\frac{V}{V^*}\bar
y^*(k_2^*-k^*_3)\bm{\delta A} -\frac{V}{V^*}\bar
y^*(k_2^*-k^*_3)\bm{\delta \chi} +
\frac{a}{V^*}\bm{\delta y}- (k^*_{-2} + k^*_2 - k^*_3)\bar y^* \,\bm{\delta \chi^*}\nonumber \\
&+& \Big( (k_2^*-k_3^*)(\frac{C}{V^*}-\frac{V}{V^*}(\bar A+\bar
\chi)-\bar \chi^*) -k^*_{-1}- k^*_{-2}\bar \chi^* -\frac{a}{V^*}
\Big)\bm{\delta y^*}.
\end{eqnarray}

Specializing to the racemic fixed point $\bar y = \bar y^* = 0$
leads to the decoupling of the first three fluctuations from the
latter two:
\begin{eqnarray}
\bm{\dot{\delta A}} &=& \big(-2k_1-[k_2+k_3]\bar \chi
-\frac{a}{V}-\frac{a}{V^*}\big)\bm{\delta A} + \Big(-(k_2+k_3)\bar
A+(k_{-2}+k_{-3})\bar \chi+k_{-1}-\frac{a}{V^*}\Big)\bm{\delta
\chi}- \frac{a}{V}\bm{\delta \chi^*}
\end{eqnarray}

\begin{eqnarray}
\bm{\dot{\delta \chi}} &=& (2k_1+[k_2+k_3]\bar \chi)\bm{\delta A} +
\Big((k_2+k_3)\bar A-k_{-1}-(k_{-2}+k_{-3})\bar
\chi-\frac{a}{V}\Big)\bm{\delta \chi}+ \frac{a}{V}\bm{\delta \chi^*}
\end{eqnarray}

\begin{eqnarray}
\bm{\dot{\delta y}} &=& +\Big(-k_{-1}-k_{-2}\bar \chi +
(k_2-k_3)\bar A-\frac{a}{V}\Big)\bm{\delta y} +\frac{a}{V}\bm{\delta
y^*}
\end{eqnarray}

\begin{eqnarray}
\bm{\dot{\delta \chi^*}} &=& -\frac{V}{V^*}(2k^*_1+[k_2^*+k_3^*]\bar
\chi^*)\bm{\delta A}
+\Big(\frac{a}{V^*}-\frac{V}{V^*}(2k^*_1+[k_2^*+k_3^*]\bar
\chi^*\Big)\bm{\delta
\chi}\nonumber \\
&+&\Big([k_2^*+k_3^*](\frac{C}{V^*}-\frac{V}{V^*}(\bar A+\bar
\chi)-\bar \chi^*)-k^*_{-1}-(k^*_{-2}+k^*_{-3})\bar \chi^* -
\frac{a}{V^*} -(2k^*_1 +
[k_2^*+k_3^*]\bar \chi^*)\Big)\bm{\delta \chi^*}\nonumber\\
\end{eqnarray}

\begin{eqnarray}
\bm{\dot{\delta y^*}} &=&  + \frac{a}{V^*}\bm{\delta y} + \Big(
(k_2^*-k_3^*)(\frac{C}{V^*}-\frac{V}{V^*}(\bar A+\bar \chi)-\bar
\chi^*) -k^*_{-1}-k^*_{-2}\bar \chi^* -\frac{a}{V^*} \Big)\bm{\delta
y^*},
\end{eqnarray}

as reflected by the specific matrix entries in Eqs.
(\ref{A33},\ref{B22}).

%-------------------------------------------------------------------
\section{\label{sec:RH} Characteristic polynomial}
%------------------------------------------------------------------

The characteristic polynomial associated with the Jacobian matrix
Eq. (\ref{Jacobian1}) evaluated at the racemic fixed point is
\begin{eqnarray}\label{charpoly5}
P(\lambda) &=& \det | J - \lambda I| = \det|A - \lambda I| \det|B -
\lambda I|,\nonumber \\
&=& P^{(3)}(\lambda)P^{(2)}(\lambda),
\end{eqnarray}
where $I$ is the identity matrix and the quadratic and cubic
polynomials are given by
\begin{eqnarray}\label{quadpoly}
P^{(2)}(\lambda) &=& \lambda^2 - {\rm
tr}\big(B\big)\lambda + \det \big(B\big),\\
\label{cubicpoly} P^{(3)}(\lambda) &=& -\lambda^3 + {\rm tr}\big( A
\big)\lambda^2 + G\big( A \big)\lambda + \det\big( A \big),
\end{eqnarray}
respectively, where
\begin{equation}
G\big( A \big) = -a_{11}a_{22} - a_{11}a_{33} - a_{22}a_{33} +
a_{32}a_{23} + a_{12}a_{21} + a_{13}a_{31}.
\end{equation}
Thus, inserting Eqs. (\ref{quadpoly},\ref{cubicpoly}) into Eq.
(\ref{charpoly5}), we obtain the fifth order polynomial
\begin{eqnarray}\label{charpolyfull}
P(\lambda) &=& -\lambda^5 + [{\rm tr}\big(B\big) + {\rm
tr}\big(A\big)] \lambda^4 + \{ G\big(A\big) - \det\big(B\big) - {\rm
tr}\big(B\big){\rm tr}\big(A\big) \}\lambda^3
\nonumber \\
&+& \{ {\rm tr}\big(A\big)\det\big(B\big)-G\big(A\big){\rm
tr}\big(B\big) + \det\big( A \big) \}\lambda^2 + \{G\big(A\big)\det
\big(B\big) - \det\big( A \big){\rm tr}\big(B\big)\}\lambda
\nonumber
\\
&+& \det\big(B \big)\det\big( A \big).
\end{eqnarray}

\footnotesize{
\bibliography{LES} %your .bib file
\bibliographystyle{rsc} %the RSC's .bst file
}

\end{document}